# Analysis of pangolin metagenomic datasets reveals significant contamination, raising concerns for pangolin CoV host attribution

Adrian Jones[1], Daoyu Zhang[2], Yuri Deigin[3]* and Steven C. Quay[4]

[1] Independent Bioinformatics Researcher, Melbourne, Australia
[2] Independent Genetics Researcher, Sydney, Australia
[3] Youthereum Genetics Inc., Toronto, Ontario, Canada; ORCID 0000-0002-3397-5811
[4] Atossa Therapeutics, Inc., Seattle, WA USA; ORCID 0000-0002-0363-7651
*Correspondence to: ydeigin@protonmail.com

## Abstract

Metagenomic datasets from pangolin tissue specimens have previously yielded SARS-related coronaviruses which show high homology in their receptor binding domain to SARS-CoV-2, suggesting a potential zoonotic source for this feature of the human virus, possibly via recombination (Liu et al. 2019, Lam et al. 2020, Xiao et al. 2020, Liu et al. 2020). Here we re-examine these published datasets. We report that only a few pangolin samples were found to contain coronavirus reads, and even then in low abundance, while other non-pangolin hosted viruses were present in higher abundance. We also discovered extensive contamination with human, rodent, and other mammalian gene sequences, which was a surprising finding. Furthermore, we uncovered a number of pangolin CoV sequences embedded in standard laboratory cloning vectors, which suggests the pangolin specimens could have been contaminated with sequences derived from synthetic biology experiments. Finally, we discover a third pangolin dataset (He et al. 2022) with low levels of SARSr-CoV sequences and unambiguous extensive contamination of several pangolin samples. For these reasons, we find it unlikely that the pangolins in question had a coronavirus infection while alive, and all current versions of the cited papers claiming a zoonotic infection of pangolins with a SARS-r CoV require substantial corrections and should be retracted until such corrections are made.

## Introduction

Pangolins, specifically *Manis javanica*, have been proposed as an intermediate host for SARS-CoV-2 by multiple authors because coronaviruses they harbored had a near-identical receptor binding domain (RBD) to that of SARS-CoV-2 (Xiao et al. 2020; Lam et al. 2020; Liu et al. 2020; Zhang, T. et al. 2020). However, the pangolin coronavirus genome sequences with an RBD that is 97% identical at the amino acid level to that of SARS-CoV-2 all stem from a single batch of smuggled pangolins (Chan and Zhan 2020). Researchers have raised the possibility that these pangolins may have been infected by humans (Choo et al., 2020; Wenzel 2020) or another host species (Chan and Zhan, 2020) during trafficking. Alternatively the viral sequences may have arisen through cell culture or other contamination during sequencing. Deigin and Segreto (2021) note the poor documentation of sample source and usage by Liu et al. (2020) and Xiao et al. (2020) in the derivation of effectively the same pangolin CoV genome (MP789 and GD_1 genome sequences respectively), while Hassanain (2020), Chan and Zhan (2020) and Zhang D. (2020) note undisclosed source documentation, irregularities and issues with the metagenomic datasets. A complex relationship of undisclosed data sharing and authorship of pangolin CoV datasets was documented by Chan and Zhan (2020), with key dates shown in Supp. Info. 0.1.

Here we review the metagenomic datasets from Liu et al. (2019), Lam et al. (2020), Xiao et al. (2020), Liu et al. (2020) and Li HM. et al. (2020) and find serious issues that raise questions as to the validity of the argument that pangolins were infected by a natural SARS-r CoV. This has consequences for SARS-CoV-2 origin analyses as multiple papers have relied on the published pangolin CoV genome sequences GD_1 (Xiao et al. 2020) or MP789 (Liu et al. 2020) or the datasets above for their interpretation (Andersen et al. 2020; Li X. et al. 2020b; Niu S. et al., 2021, Makarenkov et al. 2021). We further identify a partial sequence of a novel bat-SL-CoVZC45-related CoV in 5 *Manis javanica*, and 1 *Manis pentadactyla* pangolin and 3 *Hystrix brachyura* (Malayan porcupine) samples sequenced by He et al. (2022), but the presence of this SARSr-CoV was not detected by the authors.

Xiao et al. (2020) and Liu et al. (2020) relied on Liu et al. (2019) datasets for their analyses (Supp. Fig. S1). Other pangolin studies have published virus genomes but either relied solely on datasets from Liu et al. (2019) or Xiao et al. (2020) or have not published any metagenomic datasets for independent verification. Zhang, T. et al. (2020) relied solely on datasets (samples Lung08 and Lung07) generated by Liu et al. (2019). Yang et al. (2021) published a novel murine respirovirus strain from pangolin samples published by Xiao et al. (2020). Gao et al. (2020) identified novel pangolin pestivirus and coltivirus genomes without providing any raw sequencing data.

Li X. et al. (2020a) documented analyses of 15 Malayan pangolins confiscated from smugglers within the Guangdong Province seemingly infected with a SARS-CoV-2-related coronavirus. Those pangolin samples were collected in March-August 2019 by the Guangzhou wildlife rescue center. Li X. et al. (2020a) state that the raw sequencing data was submitted to NCBI as accession number PRJNA640246, however this BioProject could not be found on NCBI as of the writing of this paper. In Yang et al. (2021), twelve coauthors of Li X. et al. (2020a) subsequently identified a novel Murine respirovirus 'Pangolin respirovirus M5' (MW505906.1) from pangolin sample M5 from the same batch of smuggled pangolins used for the Li X. et al. (2020a) paper, but curiously failed to mention Li X. et al. (2020a).

A correction has been provided in Liu et al. (2021) that addresses issues in Liu et al. (2020) raised by Chan and Zhan (2020). Also, a correction by Xiao et al. (2021) addressed several issues identified by Chan and Zhan (2020) including previously published pangolin sample renaming and multiple data source and analysis anomalies. However, multiple issues still remain outstanding in works by Liu et al. (2019), Lam et al. (2020), Xiao et al. (2020), Liu et al. (2020) and Li HM. et al. (2020), and these issues are discussed herein.

Issues with coronavirus contamination of NGS datasets are highlighted by Katz et al. (2021) who identified more than 2000 Public Health England bacterial NGS submissions to NCBI in early 2020 likely contaminated with SARS-CoV-2 sequences. Virus contamination issues involving sequencing projects in China are also documented by Zhang et al. (2021), Quay et al. (2021a, 2021b) and Csabai et al. (2022).

In this paper we make six key observations:

1. The pangolin specimens discussed herein contain frequent and abundant contamination by sequences from a variety of other potential host species, including human and rodent sequences. This makes it impossible to conclusively ascertain the viral host in most cases.
2. The few specimens with coronavirus sequences contain evidence of numerous other viruses, many in much higher abundance. This makes it impossible to ascertain that the pangolins died from a coronavirus infection, or were even infected by it while still alive.
3. Many specimens contain large contigs of common laboratory cloning vectors with numerous viral sequences embedded, documenting contamination at some stage(s) of sequencing pipelines.
4. Pangolin coronavirus sequences are embedded in synthetic cloning vectors in the GD_1 amplicon dataset, described as a lung sample by Xiao et al. (2020) without which the GD_1 genome cannot be assembled.
5. The pangolin genomes GD_1 and MP789 cannot be assembled using the datasets specified in author corrections by Xiao et al. (2021) and Liu et al. (2021), with 15 and 3 nucleotides of their genomes not accounted for in the SRA files, respectively.
6. The discovery of SARSr-CoV reads covering the RdRp region and part of the ORF1a coding region showing highest similarities to bat-SL-CoVZC45 and pangolin-CoV GX/P4L in human sequence contaminated pangolin and rodent datasets is a further example of likely laboratory contamination leading to the presence of SARSr-CoV sequences in mammalian datasets.

For these reasons, all current versions of the cited papers claiming a zoonotic infection of pangolins with a SARS-r CoV (Liu et al. 2019, Lam et al. 2020, Xiao et al. 2020, Liu et al. 2020) must be considered unreliable, and either substantial corrections and clarifications should be provided, until which time the papers in question should be retracted.

# Results

For comparative microbial analysis of pangolin samples across multiple BioProjects, fastv was used to generate microbial profiles for pangolin metagenomic datasets published by Liu et al. (2019), Lam et al. (2020), Xiao et al. (2020), Liu et al. (2020) (documented in Liu et al. 2021) and Li HM. et al. (2020) (Supp. Info. 1.1, 2.1, 3.1, 4.1, 5.1). We identified the most abundant viruses by count and those with highest genome coverage by reads and selected 49 viruses for dataset comparison.

We then aligned each SRA to a concatenated reference set of the 49 viruses (Fig. 1, Supp Fig. S2-4). We considered a virus present in an SRA if 10% or more of the virus sequence was covered by mapped reads. This rule was relaxed for African Swine Fever Virus which was found in high abundance in vectorized format in BioProject PRJNA607174 by Xiao et al. (2020) with maximum coverage in sample M10 at 5.6%, and HIV-1, which was found to form part of a lentiviral vector with maximum coverage of 9% in MJS5 in BioProject PRJNA610466 by Li HM. et al. (2020) and included these viruses in the analyses.

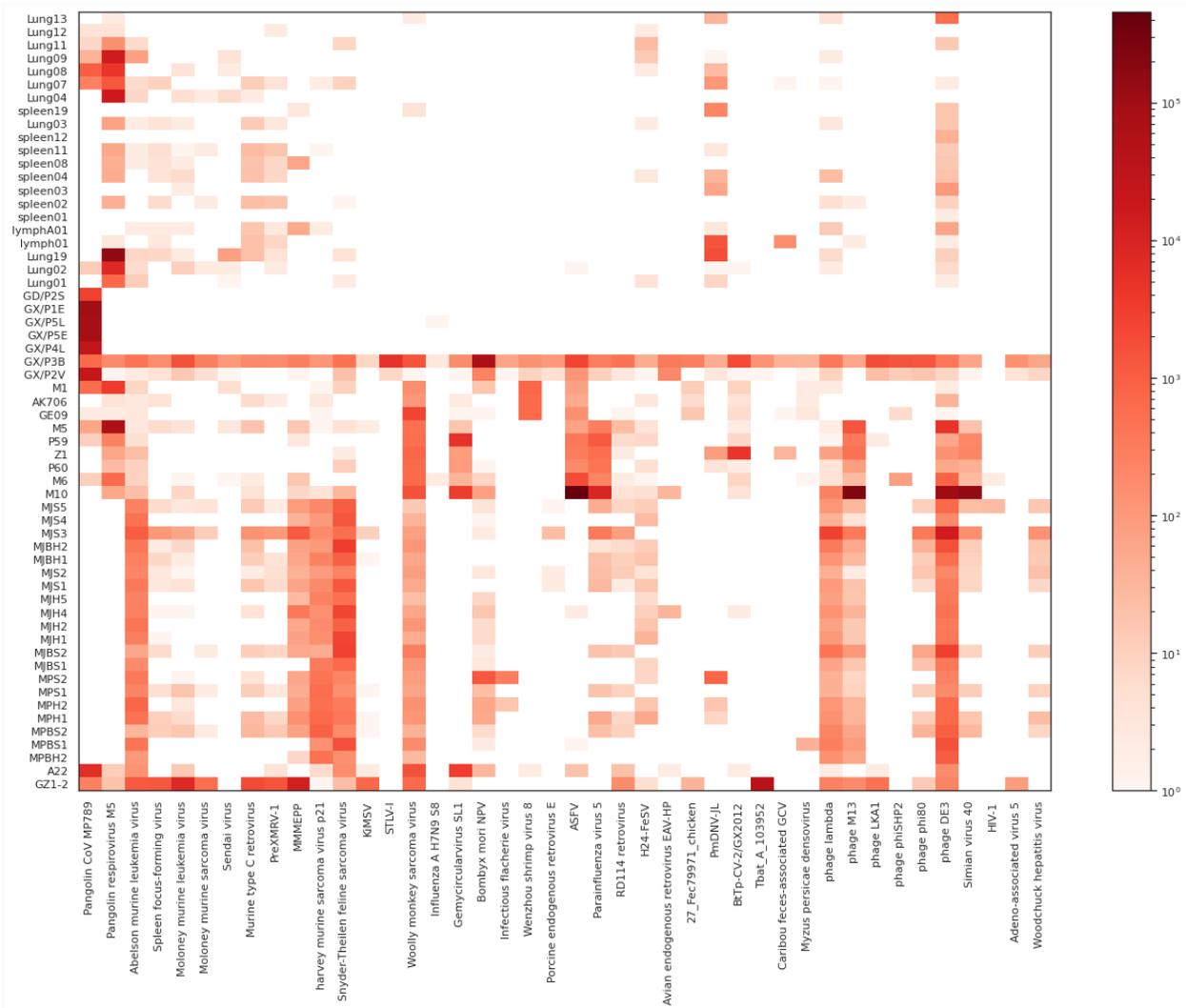

Fig. 1. Virus sequence read counts (log scale) for samples in Liu et al. (2019) (PRJNA573298), Lam et al. (2020) (PRJNA606875), Xiao et al. (2020) (PRJNA607174), Li et al. (2020) (PRJNA610466) and Liu et al. (2020) (PRJNA686836). Reads aligned using bowtie2. See Supp. Info. 0.2 for dataset information.

We de novo assembled contigs for each SRA using MEGAHIT and aligned the contigs to the same 49 viruses (Supp Figs. S5, S6) and found that 37 viruses were mapped by contigs of 141 k-mer length. Across this set of 5 BioProjects, the highest virus coverage in any single SRA was found for Pangolin respirovirus M5, Pangolin CoV MP789, Mus musculus mobilized endogenous polytropic provirus, Gemycircularvirus SL1, Wenshou Shrimp Virus 8, Parainfluenza virus 5, Parus major densovirus isolate PmDNV-JL and Bat associated cyclovirus 9 isolate BtTp-CV-2/GX2012.

Separately, we analyzed 6 pangolin and 3 porcupine samples sequenced by He et al. (2022), and identified novel bat-SL-CoVZC45-related sarbacovirus sequences, with a low read count on the order of that seen for several Liu et al. 2019 pangolin samples. Human orthorubulavirus 2

(HPIV-2) sequences were found in all pangolin samples containing ZC45-related CoV reads, with HPIV-2 read counts at an exceedingly high level in two samples, clearly indicating a contamination issue (Supp Fig. S7, S8, Supp. Info. 6.0).

We undertook a more detailed analysis of each dataset (Fig. 2.) and provide the results of our analyses under each BioProject below.

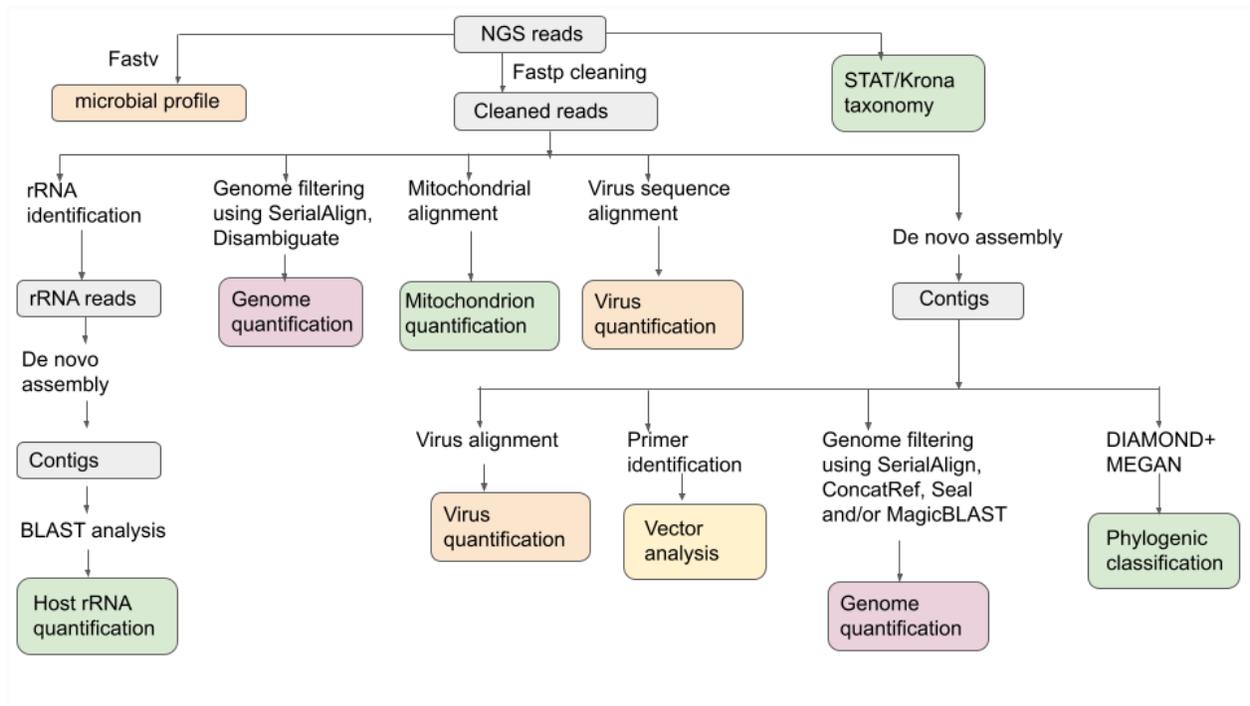

Fig. 2. NGS dataset analysis workflow.

### Liu et al. 2019 (BioProject PRJNA573298)

*Virus taxonomy*

Fastv (Chen S. et al. 2020) was used for k-mer analysis against the Opengene viral genome k-mer collection for all SRAs in PRJNA573298 (Supp. Info. 1.1). Cleaned reads in each SRA were aligned to the 49 viral sequences discussed above using both bwa mem and bowtie2, with 12 viruses having 10% or greater coverage in any SRA in PRJNA573298 (Fig. 3).

Pangolin coronavirus reads were identified in only 6 of 21 whole-genome sequencing (WGS) SRA datasets in BioProject PRJNA573298 (Fig. 1, Supp. Info. 1.3) with 76% of the reads being found in Lung08, 20% in Lung07 and 3% in Lung09.

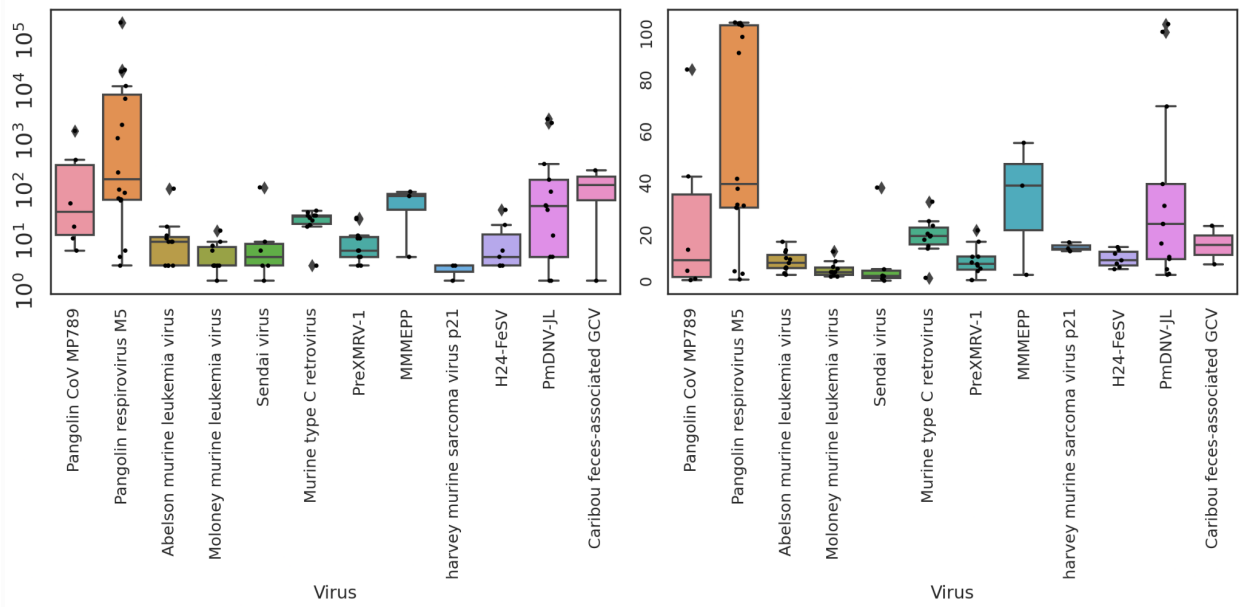

Fig 3. Distribution of total read counts and percent coverage for PRJNA573298 datasets aligned to 12 viruses with 10% or more coverage by reads in any single SRA. Read counts were identified by aligning each dataset using bowtie2. Read counts Y-axis is log scale, the percentage plot Y-axis is in percent. Box extends from the lower to upper quartile values, a horizontal line delineates the median. Whiskers extend to 1.5X of the interquartile range of the data. Outliers are drawn as diamonds. See Supp. Info. 0.4 for full titles and accession numbers.

To calculate coverage statistics across the BioProject, all 21 SRAs in PRJNA573298 were pooled and aligned to the 49 virus reference. Pangolin CoV MP789 had 1693 mapped reads with 88.3% coverage (Supp. Fig. S9. Supp Info 1.4).

The maximum read abundance of pangolin CoV MP789 (MT121216.1) in any sample is slightly lower than for Parus major densovirus isolate PmDNV-JL (NC_031450.1) but significantly lower than Pangolin respirovirus isolate M5 (MW505906.1), while median read coverages for Murine type C retrovirus (NC_001702.1) and Mus musculus mobilized endogenous polytropic provirus (NC_029853.1) are greater than for pangolin CoV MP789 (Fig. 1., Fig. 3, Supp. Info. 1.5).

*Taxonomic binning, Mitochondrion and rRNA identification*

NCBI SRA Taxonomy Analysis Tool (STAT) classification indicates significant levels of *Homininae* and *Muroidea* contamination of the SRA datasets (Supp. Info. 1.0).

To help ascertain non-pangolin species contamination, reads were mapped to a set of mammal mitochondrial genomes allowing only 100% identical matches and filtering out matchers shorter than 100nt (Supp. Info. 1.6). *Homo sapiens* mitochondrial reads were found in all samples, with highest presence in Lung02, Lung03, Lung04 and Lung08 (17%, 19%, 32% and 14% respectively). *Mus musculus* mitochondrial reads were found in 12 datasets, with highest percentages in Lung03 (25%) and LymphA01 (9%). Consistent with the findings of Hassanin (2020) *Panthera tigris* mitochondrial sequences were also found, with the highest percentage of total mitochondria in Lung09 (20%). We then aligned pooled SRA datasets allowing only 100% identical matches and calculated project level read coverages (Supp. Data. 1.7). High coverage for *Mus musculus* (94%) and *Homo sapiens* (89%) mitochondrion was found (Supp. Figs. S10-11), with moderate coverage of the *Panthera tigris* mitochondrion sequence (56%). Coverage for *Manis pentadactyla* mitochondrial sequence was 10%. A Pearson correlation matrix plot of virus and mitochondrion counts across each SRA in the BioProject shows that for Pangolin CoV MP789, a moderate correlation is found to human mitochondrial counts, and weak anticorrelation with pangolin mitochondrion counts. A Spearman correlation matrix also indicated anti-correlation to pangolin species but only a weak correlation to human and moderate correlation to mouse mitochondrion (Supp. Figs. S12-13).

Human matching rRNA of alignments lengths >150nt were identified in Lung13, Spleen11, Spleen04, Spleen01, LymphA01 and Lung02 (Supp. Info. 1.8, 1.9). We note 6.5% of the rRNA matching contigs belong to Animalia and align equally well to multiple species, but do not match Pangolin rRNA in the top 100 BLAST results. We further note Mycoplasma matching sequences, a common cell-culture contaminant (Mahmood and Ali, 2017), in Lung12, Lung11, Lung08, Lung07, Lung03, Lung19, Lung02, Lung01 and at an overall level of 2% of the rRNA matching contigs for the BioProject.

Independent taxonomic classification of *de novo* assembled contigs was generated using a DIAMOND and MEGAN taxonomic and functional classification pipeline (Bağcı et al. 2021). Order level abundance from the magnorder Boreoeutheria was plotted for comparison (Supp Fig S14). Primate matching sequences were most abundant in Lung09 and Lung01, while Rodentia matching sequences were most abundant in Spleen11 and Spleen04 (73%, 37%, 40% and 37.5% of Boreoeutheria respectively). Although up to 10% of Boreoeutheria classified contigs were classified as Chiroptera sequences (Supp. Info. 1.10), Chiroptera were not detected above trace level in mitochondria or rRNA analysis of reads. At a species level, with species with >20 contig matches, *Homo sapiens* and *Mus musculus* were identified in multiple SRA's (Supp Fig S15).

*Contamination Filtering*

To identify overall levels of human and mouse genome sequence contamination in BioProject PRJNA573298 we used SerialAlign (Fig. 4, Supp. Fig. 16) and Disambiguate (Jo et al., 2019) (Supp. Fig. 17) workflows for read analysis.

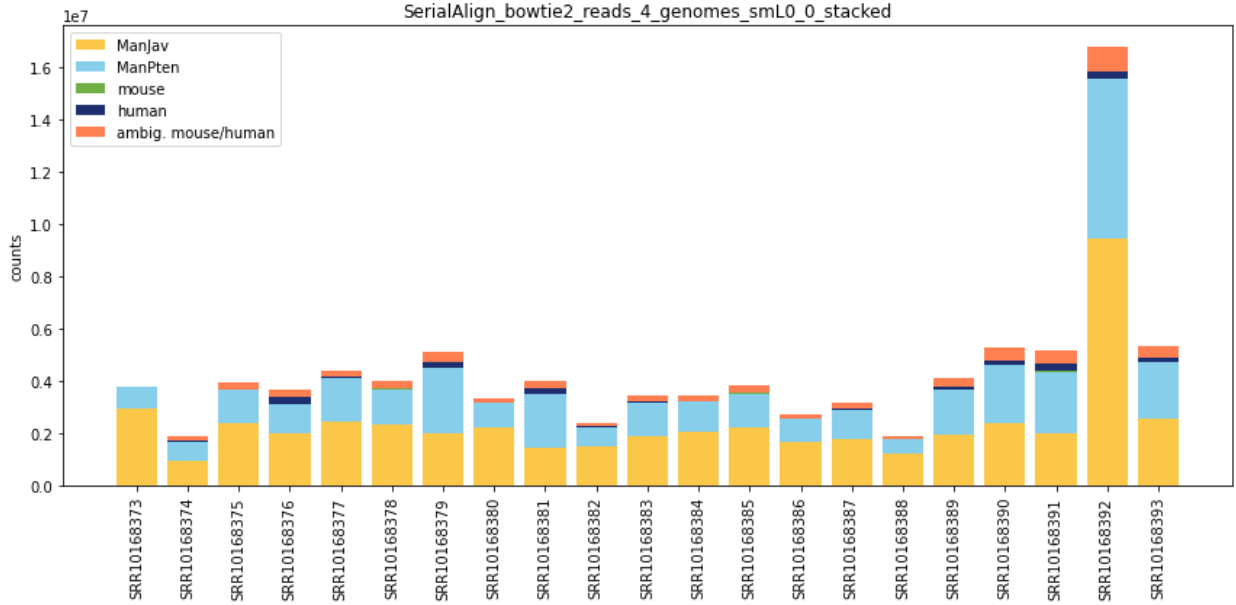

Fig. 4. Serial alignment --score-min L,0,0. Serial alignment to *Manis javanica*, *Manis pentadactyla*, human and mouse using the --score-min L,0,0 parameter. See methods for methodology details.

We then applied SerialAlign, ConcatRef (Jo et al., 2019), Seal (Bushnell, 2021) and Magic-BLAST (Boratyn et al. 2018) based workflows to contig analysis (Supp. Fig. 18-21). As contigs will likely have varying read depths, absolute match counts may not be as representative as read counts, however the longer sequences allow higher confidence alignment matches. The results of a Seal workflow using the parameter "ambig=toss" is shown in Fig. 5.

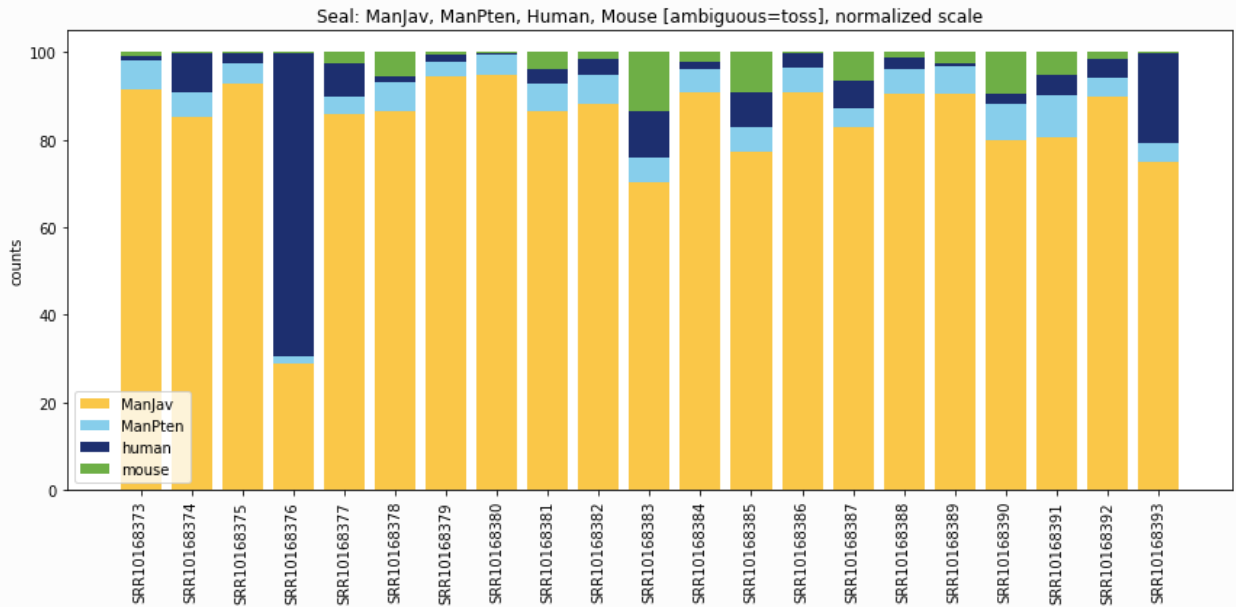

Fig. 5. Liu seal toss. Seal (BBMap suite) result for *de novo* assembled contigs mapped to a concatenated reference consisting of the YNU_ManJav_2.0, YNU_ManPten_2.0, GRCh38 and GRCm39 genomes. 'ambiguous=toss' parameter setting was used. Counts normalized to 100%.

For all 6 contig filtering workflows and the DIAMOND+MEGAN classification workflow, Lung09 and Lung01 had highest human genome matching contig counts. While highest counts for contigs matching a mouse genome were most abundant in Spleen11 and Spleen04. However for mitochondrial read alignment and read filtering using SerialAlign with 100% identity, Lung09, Lung08, Lung07, Lung04, Lung03, LymphA01, Lymph01 and Lung02 exhibited anomalously high percentages of human and or mouse matching sequences.

*Contamination with synthetic Vectors, Yamanaka factor sequences, and Protein A coding sequences found*

Fifteen contigs with common synthetic primer and/or prompter sequences as synthetic vectors were analyzed using local BLAST against the nt database (Supp. Info. 1.11). We identified a human POU class 5 homeobox 1 (POU5F1), gene sequence with a 679/683nt match to accession NM_002701.6 attached to a 188nt retroviral vector (Supp. Fig. S22). Blastn on the vector sequence identified multiple retroviral cloning and expression vectors with high homology, with gene trap vectors with accessions AB609714.1 and AB609713.1 having highest maximum score (Supp Figs. 23-25). To identify read depth and coverage distribution, we aligned pooled reads from all SRA's in PRJNA573298 to this synthetic contig using bowtie2 with default settings, and using the '--local' setting (Supp. Fig. S26). Reads contributing to the contig were found in lung02, lung03 and lung04. Noting that OCT4 is a Yamanaka factor that may be used for induced pluripotent stem cell (iPSC) growth and preparation (Takahashi and Yamanaka, 2006), we concatenated and aligned all reads to the following Yamanaka factor sequences: human SOX2 (NM_003106.4), MYC (NM_001354870.1), KLF4 (NM_004235.6) and five transcript variants for POU5F1 (NM_002701.6, NM_203289.6, NM_001173531.3, NM_001285986.2 and NM_001285987.1). We then *de novo* assembled the resulting alignments which generated 15 contigs, and used blastn to analyze alignments. 100% matches to human POU5F1 (759nt), SOX2 (888nt), and MYC (1431nt and 615nt) were found (Supp. Info. 1.12). All human iPSC matching contigs had lower percentage identity and coverage for *Manis Javanica* iPSC's.

We identified three contigs with synthetic vectors attached to Protein A coding sequences (Supp Info. 1.13). We note that protein A can be used for the purification of polyclonal antibody produced from animal antisera or monoclonal antibody from hybridoma cell culture (Hernández et al. 2020) but the source of these sequences in PRJNA573298 is unknown.

Noting that Zhang D. (2020) had identified pcDNA3.1_+ in BioProject PRJNA607174, we aligned pooled SRA datasets to this vector using bwa mem with default parameters, then extracted only reads of 150nt length with a maximum of a 1nt identity mismatch and only at read ends. We found 207 reads mapped to MN996867.1, generating a 58.2% coverage of the vector sequence (Supp. Info. 1.4, Supp. Fig. 27).

**Lam et al. 2020 (BioProject PRJNA606875)**

*SARS-CoV-2 contamination found*
Lam et al. (2020) identified pangolin-CoV GD/P2S (EPI_ISL_410544) from a Guangdong pangolin scale sampled in 2019 and 6 Guanxi pangolin coronaviruses from samples collected in 2017. The datasets by Lam et al. (2020) were registered on NCBI on 15/2/2020 and published on 26/3/2020. The SRA datasets consist of four amplicon sequencing datasets and three RNA-Seq datasets (Supp. Info. 0.2). However, the RNA-Seq dataset GD/P2S has been heavily filtered, with all non-coronavirus reads filtered out, leaving a dataset of only 2,633 reads. This compares with 55,384,421 reads for sample GZ1-2 by Liu et al. (2020), also sequenced by a NextSeq 550 instrument. We also note that sample GD/P2S was sequenced on two different machines "@NDX550382_RUO" (NextSeq 550) and "@NB501248AR" (Unknown platform) and the reads combined. Furthermore GD/P2S is the only dataset in PRJNA606875 without a description of an isolation source or a library prep. procedure.

Hassanin (2020) identified 11 reads exactly matching SARS-CoV-2 and four reads with only a single mutation difference in the GD/P2S dataset. Here we undertook a BLASTN search of SARS-CoV-2 Wuhan-Hu-1 (NC_045512.2) against the GD/P2S dataset. We found 83 reads of 75-nt length with an identical match and 132 reads of 75-nt length with a single nucleotide difference (Supp. Fig. S28). We then aligned the GD/P2S dataset to GD_1 (EPI_ISL_410721) and SARS-CoV-2 Wuhan-Hu-1 (NC_045512.2) using bwa-mem2 and further identified 4 reads 44-53 nucleotides long with an exact match to SARS-CoV-2 Wuhan-Hu-1. One 49-nt read exactly matched SARS-CoV-2 and contained the furin cleavage site sequence 'CGGCGGGCACGT'.

In sample GX/P2V, which used Vero-E6 cell line culture for pangolin CoV isolation, after alignment to SARS-CoV-2 Wuhan-Hu-1 (NC_045512.2) we identified 20 reads of 70-95 nucleotide length (70 nucleotides was used as a minimum cutoff) with an exact match to SARS-CoV-2 (NC_045512.2) (Supp. Info. 0.3 ids 30-49), indicating SARS-CoV-2 contamination of a second dataset in this BioProject.

We aligned each of the SRA datasets to a set of 49 viruses discussed previously using bowtie2 (Supp. Fig. S29). We note short read lengths in RNA-Seq dataset GX/P3B, at an average of 52nt after fastp processing. As such, alignments on reads for this dataset are lower confidence than other datasets studied. This likely explains widespread virus matches seen for reads in GX/P3B (Fig. 1) but not contigs (Supp. Fig. S5).

Using *de novo* assembled contigs aligned contigs against the NCBI viral database sequences allowed identification of several contaminating viruses not identified using fastv read analysis (Supp. Info. 2.4), with 20% of virus aligned contigs in GX/P3B matching Mus musculus mobilized endogenous polytropic provirus (NC_029853.1) (with 55% coverage), and in GX/P5L we found a 28% coverage of Influenza A virus (A/Korea/426/1968(H2N2)) segment 7 (NC_007377.1).

For taxonomic analysis we ran a DIAMOND+MEGAN classification workflow on assembled contigs. Supp. Fig. S30 highlights the filtered nature of the datasets in this BioProject with sample GX/P3B being the only sample with more than 50 contigs in total matching Eukaryota where at least 5 contigs of a species occur in any SRA in the BioProject. We generated a PCoA plot in MEGAN using DIAMOND classified contigs as input. All amplicon datasets except GX/P4L were clustered proximally, with heavily filtered RNA-Seq dataset GD/P2S also plotting in this cluster (Supp. Fig. S31).

*Mitochondrion and rRNA identification*

We aligned each SRA to mitochondria from species identified using DIAMOND+MEGAN allowing only a 100% identity match. Only GX/P3B and GX/P2V were found to contain mitochondrial sequences, where in sample GX/P3B, 4.3% of mitochondrial aligned reads matched *Homo sapiens* and low levels of matches to multiple non pangolin species (Supp. Info 2.5).

Using *de novo* assembled reads classified as rRNA sequences using Metaxa2, for all samples except sample GX/P2V, 125 contigs were 100nt or greater in length. Of these that could be unambiguously identified as non-pangolin animal species, rRNAsequences of primate origin were most abundant (11%), although given that the method used for unambiguous identification relies on 5 or less equivalent species of the same maximum alignment score, results may be skewed. Consistent with VERO-E6 cell-culture, sample GX/P2V contained 18.5% of contigs classified as mycoplasma, and 11% as primates.

**Xiao et al. 2020 (BioProject PRJNA607174): Pangolin CoV GD_1**

Metagenomic datasets were released by Xiao et al. (2020) in 4 stages (Supp. Info 0.1). WGS datasets specified in Xiao et al. (2020) that were used for generating the GD_1 genome (GISAID: EPI_ISL_410721) were published on 22/4/2020. An undocumented WGS dataset (A22) was published to NCBI on 22/6/2020. Primer-based amplicon sequences (GD1) were uploaded to NCBI on 17/11/2020 and finally two WGS datasets (AK706, GE09) which were sequenced on the same Illumina machine and run as sample M1 were uploaded to NCBI on 16/6/2021 (Supp. Info. 0.2). An author correction (Xiao et al. 2021) then noted that in addition to datasets specified in Xiao et al. (2020), WGS datasets A22 and P59 were used for GD1 assembly. Suryanarayanan (2020) notes the deletion of GE09, AK709 and GD1 and later replacement on NCBI as well as an unusual pattern of SRA data updates by Xiao et al.

*Microbial signature of datasets*

Fastv (Chen et al. 2020) was used for k-mer analysis against the Opengene viral genome collection for all SRAs in PRJNA607174 (Supp. info 3.1). We aligned each SRA in the BioProject to a set of 49 viruses as previously discussed.

Similarly to the count distribution for viruses in data from Liu et al. (2019), multiple viruses have a higher mean and maximum read count than pangolin CoV MP789, and domintly contamination-related. African Swine Fever virus sequences had, in sample M10 the highest abundance of any virus analyzed, and were found to be part of synthetic vector constructs (Fig. 6).

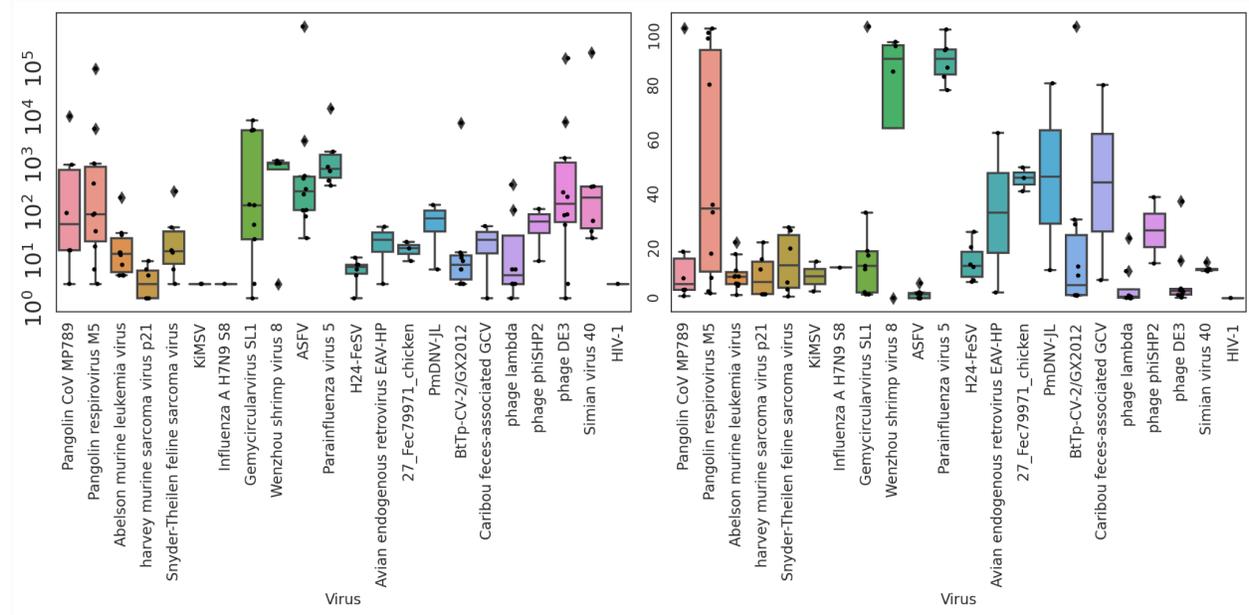

Fig. 6. Boxplot of counts and coverage, --very-sensitive. A 10% minimum coverage filter was applied to viruses, with the exception of ASFV and HIV-1. Box extends from the lower to upper quartile values, a red line delineates the median. Whiskers extend to 1.5X of the interquartile range of the data. Outliers are drawn as circles. Plotted using Matplotlib. See Supp. Info. 0.4 for full names and accession numbers.

Multiple viruses had >90% coverage (Fig. 1; Supp Figs. S2,S6; Supp. Info. 3.3). We *de novo* assembled each SRA and aligned contigs to the same set of 49 viruses. A 15015nt contig with a 97.5% coverage for parainfluenza virus 5 (NC_006430.1) was analyzed using blastn was found to have highest identity (99.94%, 14853/14862nt) to Mammalian rubulavirus 5 strain PIV5-GD18 (MG921602.1) isolated from a *Manis javanica* pangolin in March 2017 at the Guangdong Wildlife Rescue Center and suspected of causing its death (Wang et al. 2019). Parainfluenza virus 5 was found in 5 other pangolin samples, with high coverage in

PRJNA607174 all with near identical SNVs to the reference strain. Parainfluenza virus 5 was also identified in 10 pangolin samples in SRA datasets by Li HM. et al (2020) with sample MJS3 having contigs with highest coverage at 51%. The strain however is different to that seen in Xiao et al. (2020) samples, with blastn showing highest match to Parainfluenza virus 5 strain AGS (KX060176.1) (4131/4158, 1835/1841 and 1786/1794nt for 99.35% identity) isolated from human AGS cells, and a known persistent infectious agent of this cell line (Young et al. 2007). BtTp-CV-2/GX2012 was also found with high coverage in BioProject PRJNA607174, with a consensus sequence from 3 contigs assembled from sample Z1 generating 98% coverage. The consensus sequence was analyzed against the nt database using BLAST with highest identity to Cyclovirus sp. isolate CPCV isolated from the lung metagenome of a *Manis pentadactyla* pangolin sampled in May 2019 and submitted to NCBI on the 16/11/2021 by Jiangsu University (OL519623.1) (1740/1741nt for a 99.94% identity). Bat associated cyclovirus 9 isolate BtTp-CV-2/GX2012 (NC_038399.1), had the next closest identity at 96.53% (1699/1760nt). Other high coverage viruses found with contigs analyzed against the nt database for highest identity were: Caribou feces-associated GCV, with 61% coverage and 100% identity to Caribou associated gemykrogvirus 1 (NC_024909.1) in sample Z1; Parus major densovirus isolate PmDNV-JL, with 49% coverage and 97.27% identity to Densovirinae sp. isolate par081par3 (MT138268.1) again in sample Z1 and discussed in more detail later in this paper; Gemycircularvirus SL1 with 98.41% to 98% identity and in sample P59 100% coverage for multiple human hosted gemycircularvirus strains and a plant associated isolate (Supp Info. Gemy SL1); and Wenzhou shrimp virus 8 with 92% coverage and 94.29% identity to Wenzhou shrimp virus 8 strain shrimp14543 (KX883984.1) (Shi et al. 2016) found in sample AK706.

In addition to the aligned viruses in Fig 1, using minimap2 to align contigs to the NCBI viral database, we found the following viruses with >20% coverage in one or more SRA datasets: Bovine serum-associated circular virus, Mason-Pfizer monkey virus, Rhinolophus associated gemykibivirus 1, Porcine endogenous retrovirus E, Bat associated cyclovirus 11 and Classical swine fever virus (Supp. Info. 3.12).

*Taxonomic binning of reads identifies abundant non-pangolin contamination*

NCBI STAT Krona classification of reads (Supp. Info. 3.0) showed *Homo sapiens* sequences in high abundance in samples M1 and AK706 (17% and 23% respectively), with the pavorder Catarhini identified at low levels in sample A22 (0.3%). *Rattus rattus* sequences were found to comprise 16% of sample AK706 and 0.1% of sample A22, with *Mus musculus* comprising 6% of sample A22. *Penaeus vannamei* sequences were found with 3%, 5% and 0.2% abundance in samples M1, AK706 and GE09 respectively.

Due to compute limitations and the large size of some assembled datasets we used a DIAMOND and MEGAN classification workflow on *de novo* assembled contigs on only 5 of the SRA datasets in BioProject PRJNA607174. Consistent with read taxonomic classification using STAT, a high percentage of *Penaeus vannamei* (order Decapoda) contigs was apparent in samples M1, AK706 and GE09 (Fig. 1, Supp. Fig. S2). We also note trace levels of order Chiroptera sequences inferred by the DIAMOND+MEGAN workflow in samples M10, M5 and A22 (Supp. Fig. S33).

*Contamination Filtering*

Seal (Bushnell, 2021) was used for *de novo* assembled contigs alignment quantification against the *Homo sapiens*, *Mus musculus*, *Manis pentadactyla*, *Manis javanica* and *Panaeus Vannamei* genomes in two workflows (Supp. Fig. S34-35). Also broadly consistent with STAT analysis M1, AK706, GE09 contigs contained significant percentages of *Penaeus vannamei*, and variable percentages of *Homo sapiens* and *Mus musculus* contamination.

*rRNA identification and Mitochondrion sequence matching*

4.5% of contigs matching rRNA sequences unambiguously matched *Mycoplasmataceae* rRNA, the highest overall level of all BioProjects studied (samples M1, M5, P60, P59, GE09 and A22), Rodentia, Primates and *Penaeus vannamei* were found at 2.7-1.5% levels, with 1.2% of contigs matching Chiroptera rRNA (Supp. Info. 3.13, Supp. Info. 3.14).

*Penaeus vannamei* mitochondrion matching reads comprised 9.0% to 36% of matched mitochondrial sequences in AK706, M1 and GE09 (Supp. Info. 3.10), while *Homo sapiens* mitochondrion percentages were between 1.8% and 11.26% in AK706, M1, GE09 and A22 all with >55% coverage. *Mus musculus* genome coverage was 79.9% in A22.

The mitochondrion alignments are consistent with results from read taxonomic analysis using STAT, and assembled contig taxonomic analysis using DIAMOND+MEGAN, and whole genome filtering using Seal, and we infer high confidence in significant *Penaeus vannamei* and *Homo sapiens* contamination and lower levels of *Mus musculus* sequence contamination in this BioProject.

*Sample A22*

Sample A22 was submitted to NCBI on 19/6/2020, over a month after WGS dataset release on 22/4/2020 and paper publication on 7/5/2020. The sample was sequenced on a different Illumina NovaSeq 6000 machine id ("A00151") to the other WGS datasets (Supp. Info. 0.2) in this BioProject and contains the highest Pangolin CoV read count of all WGS datasets in the project. Indeed, read counts are 6.3 times the highest SARSr-CoV read count for any pangolin sample sequenced by Liu et al. (2019) (Supp. Fig. 1). Chan and Zhan (2020) noted that in Extended Data Table 3 of Xiao et al. (2020) nine pangolin lung samples were used in the study, 7 with coronavirus reads from which the GD_1 genome (GISAID: EPI_ISL_410721) was assembled from contigs. Sample A22 was not specified in the datasets used by Xiao el al. (2020). An author correction (Xiao et al. 2021) then stated that for GD_1 genome assembly, Xiao et al. (2020) used

9 pangolin samples with coronavirus sequences, including previously undeclared samples A22 and P59. However Xiao et al. (2021) did not explain why no mention of sample A22 was made in Xiao et al. (2020). We note that Xiao et al. (2020) was submitted to Nature on the 16/2/2020 and accepted on the 28/4/2020, and Xiao et al. (2021) mention that A22 was sequenced in March 2020. Another pangolin sample 'A22-2' presumably related to A22 was referenced by Li X. et al. (2020a) as a SARSr-CoV-2 infected pangolin, but no raw sequence data has been made available.

We aligned each SRA in BioProject PRJNA607174 to the Pangolin CoV MP789 genome using bwa mem. Read coverage is shown in Fig 7. Sample A22 had the best coverage of the WGS datasets.

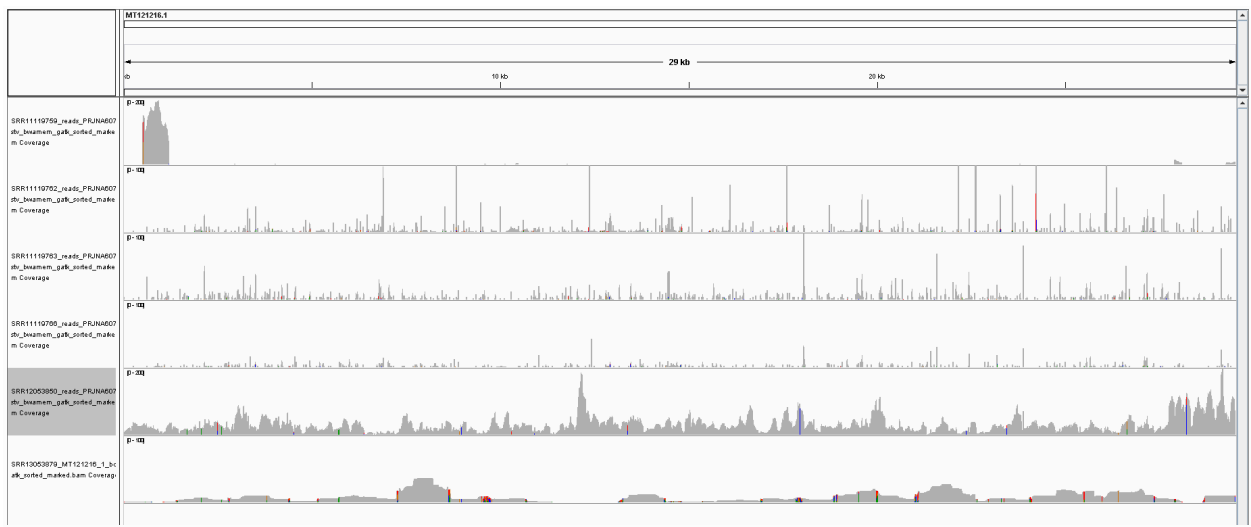

Fig. 7. Read coverage plots for Xiao et al. (2020) samples (top to bottom) M1 (0-200 scale), M5 (0-100 scale), P59 (0-100 scale), M6 (0-100 scale), A22 (0-200 scale) and GD1 amplicons (0-100 scale) aligned to MP789 (MT121216.1), plotted in IGV.

We pooled the nine pangolin samples which contained pangolin coronavirus mapped reads from Xiao et al. (2021) Author Correction: Extended Data Table 3 and aligned the combined dataset to the GD_1 genome using both bowtie2 and minimap2. We found a 48.25% coverage at 30X read depth, a recommended depth for robust genome sequencing (Hassanin, 2020; Sims et al., 2014), and 99.95% coverage total of the GD_1 genome. This compares with only a 5.39% coverage at 30X read depth, and 90.63% coverage total of the GD_1 genome using the original Xiao et al. (2020) datasets aligned to the GD_1 genome using bowtie2 (Supp. Info. 3.15).

Even with the inclusion of the A22 dataset, the complete GD_1 genome cannot be generated. A 15nt section at the 5' end of the genome is not covered by any reads in the datasets specified in Xiao et al. (2021). We also note four unanimous (i.e. across all reads in A22) single nucleotide variations in dataset A22 compared to the GD_1 genome: T22487C (7 reads), T23085A (5 reads), T23503A (14 reads), T24228C (28 reads) (Supp. Info. 3.16). Two of these can only be accounted for in the GD_1 genome by Xiao et al. (2020) by including the amplicon dataset GD1. For T23503 only one read from datasets included in Xiao et al. (2021) Author Correction: Extended Data Table 3, other than dataset A22 covers this nucleotide position (Supp. Info. 3.16). In contrast to the GD_1 genome, all 14 reads in A22 and the single read from Lung08 in the pooled datasets from Xiao et a. 2020 Ext Table 3 have an adenine at position 23503. Similarly, at position T24228, three reads from datasets included in Xiao et al. (2021), other than dataset A22 cover this nucleotide. However, these three reads and the 28 reads from dataset A22 all have a cytosine at this location. We note that the 15nt gap at the 5' end of the GD_1 genome not covered by datasets in Xiao et al. (2021), is covered by the amplicon dataset GD1 (Supp. Fig. S36). The dataset GD1 also contains consensus T23503, T24228 sequences.

*Amplicon dataset*

In the third data release by Xiao et al. (2020), an amplicon dataset GD1 was published on NCBI on 17/11/2020, after Chan and Zhan (2020) noted that data supporting gap coverage had not been provided by Xiao et al. (2020). In an author correction Xiao et al. (2021) describe the GD1 amplicon sequence as thus "The raw sequence data (including the trace files) generated by PCR for the assembly of the Pangolin-CoV genome have been deposited to the SRA database of NCBI (accession no. SRX9503273)". We checked for standard adapter sequences using FastQC and searched for the presence of the first 12/13nt of standard adapter sequences but did not detect the presence of any matches. We then ran fastp to filter the dataset with 240 of 240 reads passing filtering. We aligned the GD1 amplicon reads to the GD_1 genome (EPI_ISL_410721) using bwa mem with default parameters. The aligned amplicon reads as well as raw read coverage is shown in Fig. 7, with a zoomed in view of the 3' end shown in Supp. Fig. S37.

10 reads in the GD1 amplicon sequence dataset extended downstream past the 3' end of the GD_1 genome (Supp. Fig. 38). Read id '12' (Supp. Info. 0.3), one of the two 3'-most reads in Supp. Fig. 38 was found to contain M13 primer and lac promoter sequence (Fig. 8). The GD_1 sequence terminates at 538 nt in this read sequence.

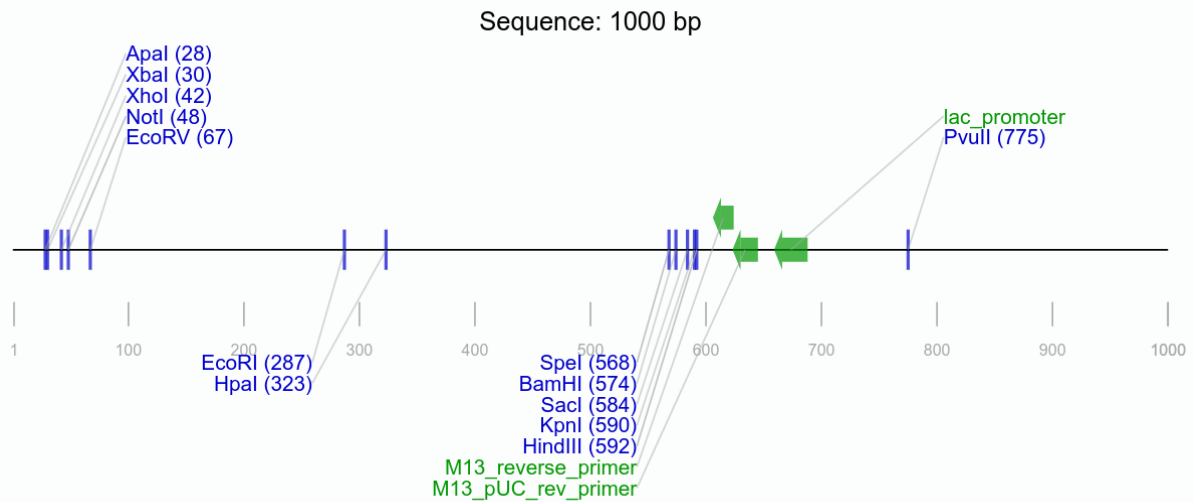

Fig. 8. Amplicon read '12' (Supp. Info. 0.3) plotted using Addgene sequence analyzer.

The part of the read sequence outside the GD_1 genome (539-1000 nt) was then analyzed using NCBI BLASTN against the nt database using default settings with highest alignment score to multiple cloning vectors (Supp. Fig. S39-40). The 81-nt misaligned section at the 5' end of the read was also analyzed using NCBI blastn, resulting in a match with a 100% identity to a 75-nt section of several pEASY cloning vectors (Supp. Figs. S41-42).

GD_1 amplicon reads aligned to the 5' end of the GD_1 genome are shown in Fig. 9 below. Notably, 13 amplicon reads extend upstream of the GD_1 genome sequence, with 9 extending over 500 nucleotides upstream. Also notable is that the amplicons cover a ~120-nt gap in pooled read coverage at the 5' end of the GD_1 genome sequence.

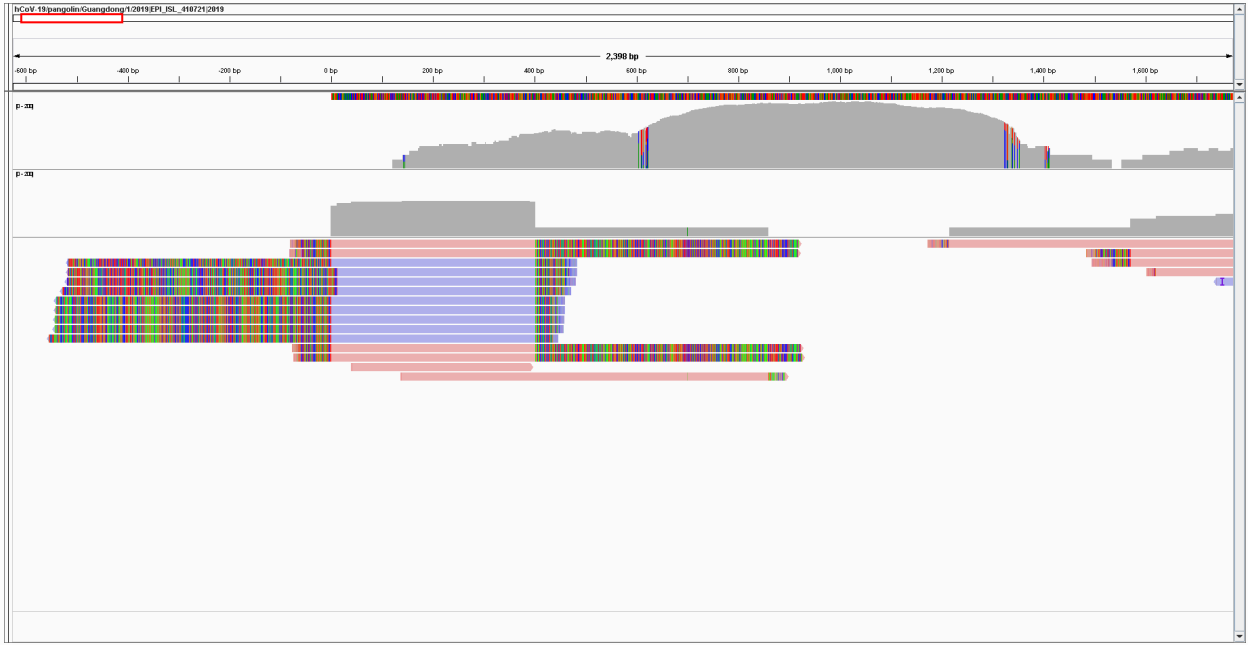

Fig. 9. Pooled reads from SRAs in Xiao et al. (2020) Extended Data Table 3 coverage aligned to GD_1 genome with bowtie2 (gray, top track, 0-200 log scale); GD_1 amplicon reads aligned to GD_1 genome with bwa mem: coverage track (grey, 0-200 log scale), and read track, coloured by read direction (pink/blue) and mismatched nucleotides (strong colors). Zoomed in to the 5' end of the GD_1 genome (-600 to 1650 nt). Read id '25' is 5' most read, read id '21' is 7th from top. Displayed in IGV.

The 5' end of read id '25' (Supp. Info. 0.3) which was outside of the GD_1 genome alignment (1-554 nt) was analyzed using blastn resulting in a match with 100% identity to pEASY cloning vectors (Supp. Fig. S43).

We then analyzed the 59-nt misaligned section of the 3' end of read id '21' (Supp. Info. 0.3) using blastn and found a 100% match for a 50-nt section of several pEASY cloning vectors. We then spliced the synthetic section of the two read ids '25' and '21' (Supp. Info. 0.3) to form a more complete synthetic vector sequence (Supp. Fig. S44). The spliced sequence was then analyzed with blastn and a 99% match (603 of 604 nucleotides) was found to cloning vector pEASY-T1 (Supp. Fig. S45-46)

We further note that a large amount of reads in the amplicon dataset GD1 exhibit a distinct pattern of misalignment to the GD_1 genome sequence at either the 5' or 3' ends or both. We found that 90% of the 240 amplicon reads had read ends where >14-nt long contiguous sections misaligned to the GD_1 genome. We used a local BLAST search against the nt database and identified that 85 of the misaligned read ends matched the cloning vector pEASY-tub/hptII, 4 matched pEASY-tub/aadA and 1 matched vector pEASY-tub/ble.

After *de novo* assembly of the SRR13053879 amplicon dataset, two contigs were identified with M13 primer sequences and found to match multiple cloning and expression vectors (Supp. Figs. S47-52).

In all, 48 amplicon reads had either 'M13R' or 'M13F' in the read name, 18 of which had either complete M13_reverse_primer, M13_pUC_rev_primer and NotI restriction enzyme sequences or complete T7_promoter and NotI restriction enzyme sequences.

*Synthetic vectors in WGS datasets*

We identified contigs with common synthetic primer/promoter sequences using a pattern matching workflow in python, then identified open reading frames and translated these to proteins and analyzed either against a local copy of the nr database or using the online NCBI nr database (Supp. Info. 3.7). We recovered protein codings for 51 translated ORF's, with 31 having sequences coding for African Swine Fever virus (ASFV) proteins. Large contigs containing a Porcine circovirus 2 capsid protein from sample M5, and ASFV proteins (from sample Z1) were found (Supp. Figs. S53-54). We note that several recent ASFV related publications by South China Agricultural University (SCAU) affiliated authors (which published BioProject PRJNA607174): *in vitro* viral RNA transcription inhibitor research (Huang et al. 2021), a vaccine overview (Wu et al. 2020), and rapid detection methodology (Bai et al. 2019). SCAU affiliated authors have also previously published Porcine circovirus 2 related research (Zhai et al. 2014; Wei et al. 2019; Zhai et al. 2017).

**Li HM. et al. 2020 (BioProject PRJNA610466)**

Li HM. et al. (2020) specify in materials and methods that fourteen skin samples of *M. javanica* animals were collected from the Guangdong Provincial Wildlife Rescue Center. However, the date when the samples were collected was not provided. The authors also did not specify if the samples were sourced from the same batch of pangolins sampled by Liu et al. (2019) or if data from BioProject PRJNA573298 (Liu et al. 2019) was used. We found that the Illumina read headers showed that the same machine id was used "A00184" (runs 448 and 449) as for the 9 SRAs from Liu et al. (2019) (runs 351 and 352) (Supp. Info. 0.2).

Although neither Pangolin CoV MP789 or Pangolin respirovirus M5 were identified, we did identify, using fastv, several virus genome sequences common to BioProject PRJNA573298 and PRJNA610466 (Supp Info 4.1). We aligned each SRA in BioProject PRJNA610466 to a set of 49 viruses as discussed previously (Fig. 1). We found 1168 reads in sample MJS3 for 91% coverage of Mus musculus mobilized endogenous polytropic provirus (NC_029853.1), 781 reads and 68% coverage of Parus major densovirus isolate PmDNV-JL (NC_031450.1) in MPS2, and 80 reads for a 65% coverage of xenotropic feline endogenous retrovirus RD-114 (NC_009889.1) in MJS3 (Fig. 10, Supp. Figs. S2, S55).

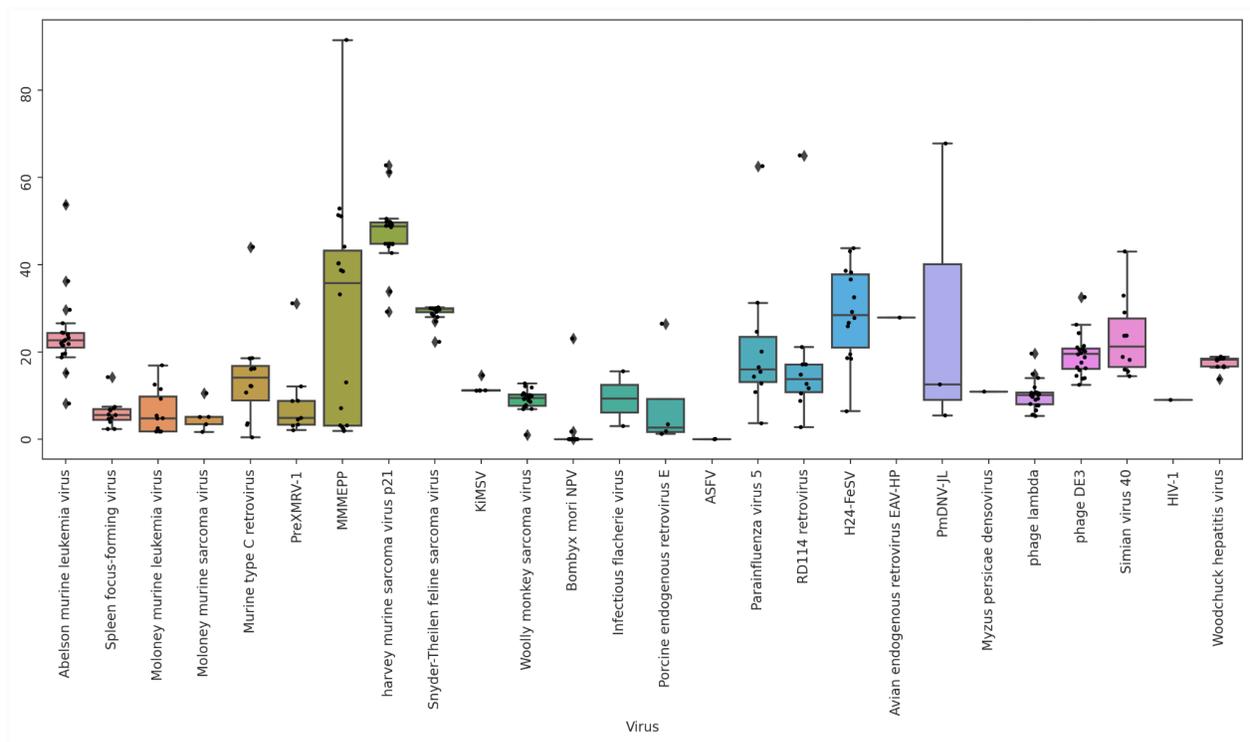

Fig. 10. Box plot of distribution of percentage coverage of all datasets in PRJNA610466 aligned to 49 viruses. Only viruses with at least one SRA having reads with >=10% coverage of the virus genome are plotted.

We aligned contigs using minimap 2 to the NCBI viral database, with abundant murine hosted viruses identified (Supp. Info. 4.4).

*Taxonomic binning*

Online NCBI SRA Taxonomy Analysis Tool (STAT) (Katz et al. 2021) was used for taxonomic classification (Supp. Info. 4.0). Sample MJS3 appeared most anomalous, with 14% bacteria compared with an average of 0.47% for other samples in the BioProject. STAT classification identified 1% *Homo sapiens* and 0.8% *Mus musculus* homologous sequences in this sample. Overall, STAT classification identified multiple samples in this BioProject containing low levels (0.1-1%) of Simiiformes, Homininae, *Homo sapiens*, *Myotis*, Chiroptera, or Rodentia matching reads.

On assembled contigs, DIAMOND+MEGAN was used to classify and plot taxonomy for Laurasiatheria with >1000 contig matches at order level. Low levels of Chiroptera, Rodentia and Primates across all SRA's except sample MJS3 where higher levels of Rodentia and Primate sequence matching contigs were classified, consistent with STAT classification of reads (Supp. Figs. S56-57).

*Mitochondrial alignment and rRNA identification*

Cleaned reads from each SRA in PRJNA610466 were aligned to a concatenated set of mammal mitochondrial genomes using bowtie2 (Supp. Info. 4.5). Over 2000 reads forming greater than 85% coverage of the *Mus musculus* mitochondrion sequence (KY018919.1) were found in 10 SRA's with 248997 matching reads in MJS3 for a 93% coverage of the *Mus musculus* mitochondrion genome at a 30X read depth. *Homo sapiens* mitochondrion matching reads were again most abundant in MJS3 with a 98% genome coverage and with over 1000 reads for a >79% genome coverage in 9 SRA's. Sus scrofa mitochondrion sequences were found at 0.34% in MPBS2.

Contigs matching rRNA sequences of single species were dominated by bacterial, *Manis pentadactyla* and *Manis javanica* rRNA. We note however single contigs with *Pipistrellus pipistrellus* genome matches of 243-267nt length in 8 SRAs (Supp. Info. 4.6).

*Synthetic vectors*

We identified an integration junction between a lentiviral vector and *Manis Pentadactyla* genomic DNA in a contig in sample MJS3 using blastn (Supp. Fig. S58-61). In the same sample MJS3 we also identified a 3123-nt contig with a 98.28% identity to and 60% coverage for Lentiviral transfer vector FUW-tetO (MK318529.1) with an open reading frame with a 100% match for 694 nt to expression vector pSOSV/ZsG-FL using blastn (Supp Figs. S62-64). In the same dataset we also identified sections of SV40 Large T antigen (Supp. Fig. S65).

**Liu et al. 2020 and Liu et al. 2021 (BioProject PRJNA686836): Pangolin CoV MP789**

The pangolin CoV MP789 genome assembly was documented by Liu et al. (2020). After Chan and Zhan (2020) noted that only two of the three pooled pangolin samples used by Liu et al. (2020) were available on NCBI and these had numerous gaps in coverage of the MP789 genome, a correction in Liu et al. (2021) was made and the sample GZ1-2 and ABIF raw chromatogram dataset 'journal_ppat_1009664_s001' were deposited to NCBI. We aligned each of the datasets documented in Liu et al. (2021) to the MP789 genome (Supp. Fig. S66) using both bwa-mem and minimap2. We observed a 3nt gap at 5716-5718nt in coverage of the pangolin CoV MP789 genome (MT121219.1) by the 4 datasets claimed by Liu et al. (2020) and Liu et al. (2021) as having been used to generate the MP789 genome. We also note a SNV at 5731C with only 2 mapped reads covering this position 5731A and 5731G both in Lung08. It seems remarkable that after issues raised by Chan and Zhan (2020), the correction by Liu et al. (2021) still did not provide all the datasets that would allow MP789 assembly.

We then analyzed the RNA-Seq dataset GZ1-2 in more detail. Using fastv for analysis, we note a microbial profile distinct from other Guangdong Institute of Applied Biological Resources (GIABR) pangolin samples (Supp. Fig. S67; Supp. Info. 5.0).

Using bowtie2 we aligned the GZ1-2 dataset to 49 viruses initially identified in the pangolin datasets using fastv (Supp. Info. 5.2), the only notable coverage percentage found was for Pacific flying fox faeces associated gemycircularvirus-1 isolate Tbat_A_103952 at 31.6% for sample GZ1-2. Contrastingly, dataset GZ1-2 has a low read count (271) and very low coverage (2.7%) for Pangolin CoV MP789 (Supp. Fig. S68). Only 16 reads had any unique coverage, with each read repeated between 2 and 44 times, indicative of high cycle PCR amplification and high number of read repeats is seen with other viruses sequenced. The virus sequenced in GZ1-2 appears to have several SNVs compared with MP789 with the following differences: T1818 (22 reads), C16543 (23 reads), T16827 (19 reads), T27123 (2 reads), T28268 and a 6nt insert "GTTGTT" between 28267 and 28268 (7 reads). No reads provided any additional MP789 genome coverage over pooled Lung07 and Lung08 samples (Liu et al. 2019), two of the three samples used by Liu et al. 2020 (Supp. Fig. S1).

We aligned de novo assembled contigs for dataset GZ1-2 to a reference sequence consisting of concatenated NCBI viral databases (Supp. Info. 5.3). Murine type C retrovirus, Pacific flying fox faeces associated gemycircularvirus-1 isolate Tbat_A_103952, and Sapovirus Mc10 were found to have the highest coverage. Coronavirus MP789 was not identified as read counts were too low to form contigs of sufficient length.

NCBI STAT analysis shows 4% of reads match *Rattus rattus*, and 3% match *Homo sapiens* species. Alignment to mitochondrial sequences indicated only low levels of *Homo sapiens* mitochondrion matching reads were found, with a 11.7% coverage (Supp. Info. 5.6), Given the short read length of 50nt, read alignment confidence is lower than for other datasets analyzed. Contig taxonomic analysis using DIAMOND and MEGAN indicated low levels of Chiroptera, Artiodactyla, Primates and Carnivora sequences (Supp. Info. 5.7).

**He et al. 2022. (BioProjects PRJNA793740 and PRJNA795267): novel bat-SL-CoVZC45-related CoV**

He et al. (2022) undertook virome characterization on 1725 game animals across China sampled between 2017 and 2021 including 21 *Manis javanica* and 12 *Manis pentadactyla* pangolins and 402 *Hystrix brachyura* (Malayan porcupines). He et al. note that pangolin samples were all obtained from Zhejiang province in eastern China, with 25 samples collected prior to February 2020, and 11 sampled between 2017 and 2019 (inclusive).

*SARSr-CoV reads*
Of the 32 pangolin samples, 24 pangolin SRA datasets were uploaded to NCBI by He et al. and we used NCBI STAT to analyze their taxonomy. SARSr-CoV matching reads were detected in six pangolin samples: MJ-ZJ-MO-1/2/3/4/6 and MP-ZJ-MO-4. Each SRA dataset was aligned to

65 viruses including viruses with significant coverage identified using fastv analysis, and of several SARSr-CoVs we found highest coverage for bat SARS-like coronavirus isolate bat-SL-CoVZC45 (MG772933.1) (Supp. Info. 6.1). Bat-SL-CoVZC45 aligned reads were merged to calculate overall coverage statistics with 429 reads mapping to the bat-SL-CoVZC45 sequence for a 9.39% coverage. A consensus sequence was then analyzed using NCBI blastn with highest maximum score, total score and percentage identity to Bat SARS-like coronavirus isolate bat-SL-CoVZC45 (MG772933.1) (2743, 4486, 94.79% respectively).

The reads from pangolin datasets are concentrated in the nonstructural protein 4 (NSP4) N-terminal and C-terminal domains in ORF1a and NSP10 and RNA polymerase (RdRp) coding regions in ORF1ab with a significant gap from 15222nt to15438nt and two single nt gaps at 14056nt and 14748nt (Supp. Fig. S69). Relatively high read depths occur over parts of the RdRp region. Of the 6 pangolin 2 samples: MJ-ZJ-MO-4 and MJ-ZJ-MO-6 had only 2 SARSr-CoV aligned reads. The four pangolins samples with >2 reads of SARSr-CoV coverage all exhibit a similar coverage pattern (Supp. Figs. S70-71). Two short regions in the ORF1a NSP4 region are covered by 5 reads, with both MJ-ZJ-MO-3 and MJ-ZJ-MO-6 containing reads in this region. The 3' end of read alignments is distinctly ended at 16239nt, with reads in MJ-ZJ-MO-1, MJ-ZJ-MO-2 and MP-ZJ-MO-4 all exactly ending at this position, which we note is not a transcription boundary. We also note 24 reads from the 6 pangolin samples better mapped to Guangxi pangolin viruses P1E/P2V/P3B/P4L/P5E/P5L including 6 reads in the sparsely covered ORF1a NSP4 coding region (Supp. Info. 7.2).

*Rodent datasets: (PRJNA795267)*
After analysis of pangolin datasets in PRJNA793740 we found He et al. (2022) had also generated a rodent meta-transcriptomic analysis BioProject PRJNA795267. We identified SARSr-CoV reads in 3 *Hystrix brachyur*a samples in this BioProject, two sourced from Fujian province and one from Hubei province (HB-FJ-NA-7, HB-HuB-N-3, HB-FJ-NA-3) in May through July 2020. While the six pangolin samples with SARSr-CoV matching reads were all sequenced on the same Illumina machine ("A00583"), run, flowcell and lane (Supp. Info. 0.2), the three Malayan porcupine samples were sequenced on two different machines, one on the same machine, run and flowcell as the six pangolin samples but on a separate flowcell lane, and two on a different Illumina machine ("A00301") with both sequenced on the same run and flowcell. We aligned each of the 3 SRA datasets to viruses identified using fastv as well as the bat-SL-CoVZC45 and Guangxi pangolin viruses identified in PRJNA793740 (Supp. Fig. S72-S73). We were surprised to find what appears to be exactly the same NSP10 and start of the RdRp coding region as the novel bat-SL-CoVZC45-related virus found in the pangolin samples (Supp. Fig. 74). 24 SNVs between 13477 and 14056nt were exactly the same in sample HB-FJ-NA-7 as for a merged alignment of 6 pangolin samples from PRJNA793740. Alignment of merged reads to bat-SL-CoVZC45 and pangolin GX_P4L generated 12.99% and 11.37% coverage respectively (Supp. Info. 6.3, Supp. Data 6_1, 6_2). Of the nine pangolin and Malayan porcupine samples, Malayan porcupine sample HB-FJ-NA-7 had 5.5X more SARSr-CoV aligned reads than pangolin sample MJ-ZJ-MO-2 which had the second highest read count (Supp. Info. 7.2).

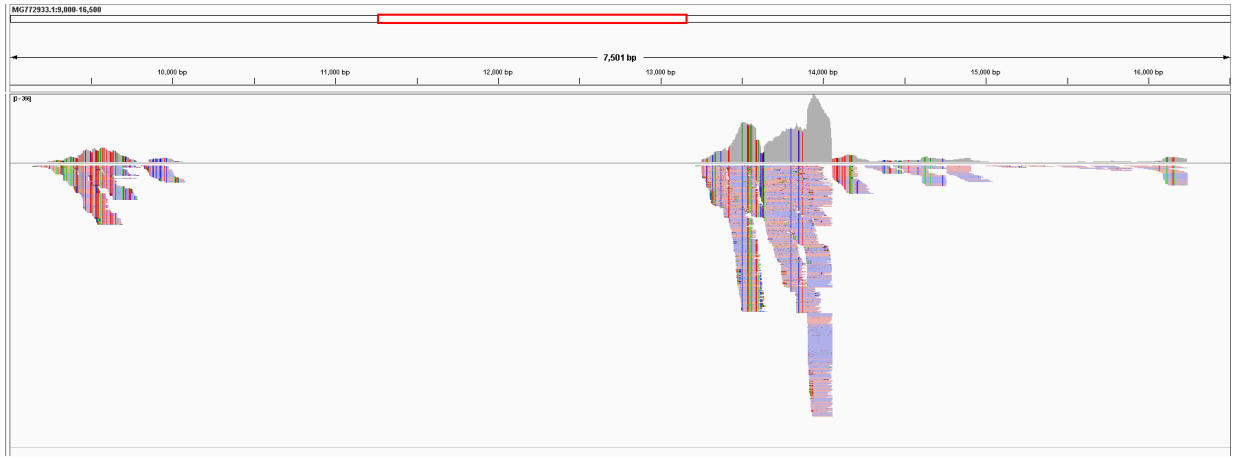

Fig. 11. Alignment of 6 pangolin datasets (MJ-ZJ-MO-1, MJ-ZJ-MO-2, MJ-ZJ-MO-3, MJ-ZJ-MO-4, MJ-ZJ-MO-6 and MP-ZJ-MO-4) and 3 porcupine datasets to bat SARS-like coronavirus isolate bat-SL-CoVZC45. Reads were aligned using bwa-mem, merged and plotted in IGV. Coverage scale 0-113. Zoomed to the 9000-16500nt region showing the complete extent of read coverage.

All reads from the 6 pangolin and 3 Hystrix brachyura SRA datasets were then pooled as per methods and *de novo* assembled. Paired-end and single-end contigs were each aligned to bat-SL-CoVZC45 and GX pangolin CoV's (Fig. 11). The addition of the 3 *Hystrix brachyura* datasets resulted in increased coverage of both a NSP4 section of ORF1a, allowing complete coverage the NSP4 N-terminal and C-terminal domains (9,137-10,077nt (relative to MG772933.1)) as well as the ORF1a NSP10 and near complete coverage of the ORF1b RdRp coding region (13,241-16239nt). Increased aligned read count and coverage was found in the NSP4, N and C termini regions when the reads were aligned to pangolin CoV GX P4L, and a better match found to the NSP10 and RdRp region when reads were aligned to bat-SL-CoVZC45 (Table 1). A more noisy SNV pattern was noted in the NSP10 and high coverage section of the RdRp (13,241-14056nt) than other mapped areas (Supp. Fig. 75).

| Reference | ORF1a Nsp4_N,C (9,137-10,077nt) | | | ORF1ab Nsp10,RdRp (13,241-16239nt) | | |
|---|---|---|---|---|---|---|
| | N | Pct. identity | Identities | N | Pct. identity | Identities |
| PCoV GX/P4L | 253 | 94.89% | 892/940nt | 892 | 92.88% | 900/969nt |
| Bat-SL-CoVZC45 | 180 | 94.13% | 593/630nt | 1064 | 95.70% | 2872/3001nt |

*Table 1. 6 pangolin and 3 Hystrix brachyura SRA datasets were pooled and aligned to bat-SL-CoVZC45 and pangolin CoV GX P4L with consensus sequence from each major aligned section analyzed using blastn and blastx.*

Simplot analysis of the mapped regions shows the NSP4 N and C termini regions with high similarity to pangolin CoV GX P4L over all but the 5' most end of this region. For the NSP10 region fairly equivalent similarity is seen to pangolin CoV GX P4L, bat-SL-CoVZC45 and four most closely related virus strains. Over the RdRp region several regions exhibit 100% identity to bat-SL-CoVZC45 (Fig. 12).

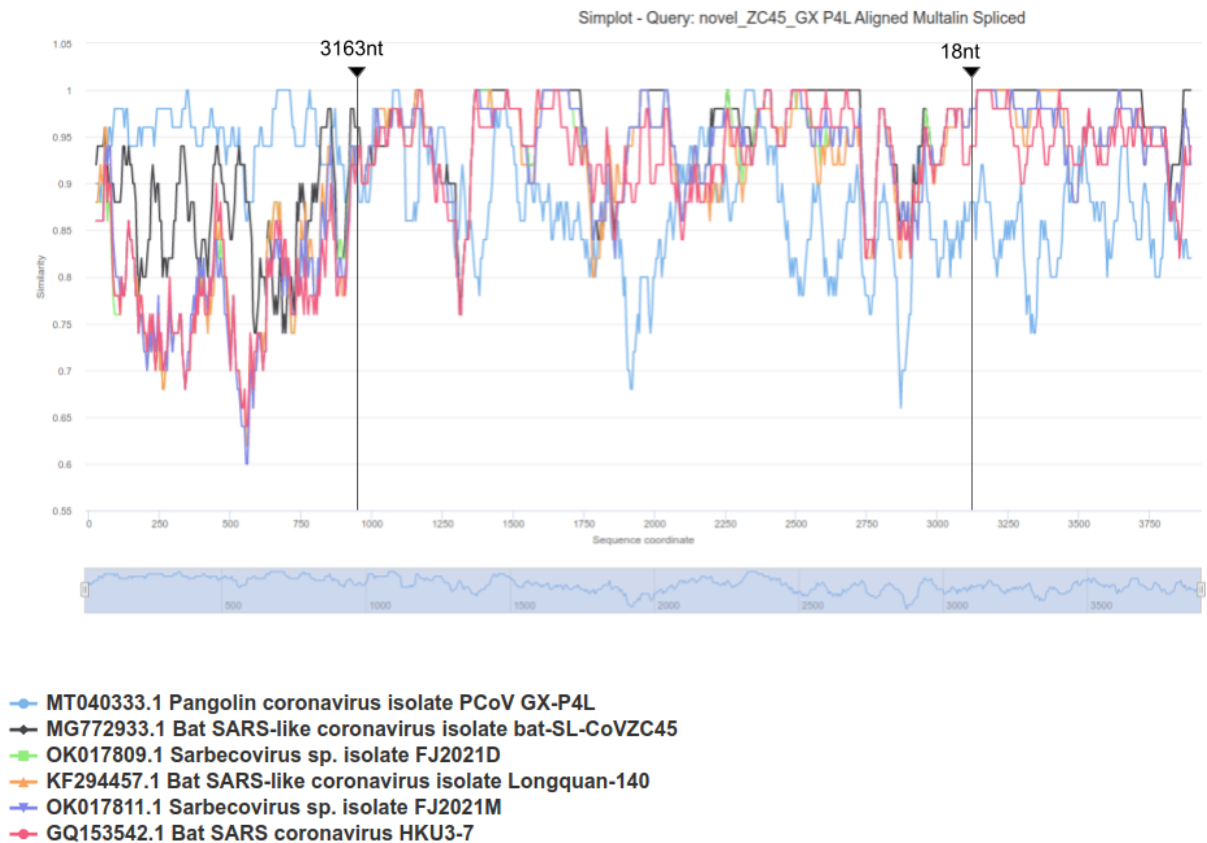

Fig. 12. Similarity plot of 5 closest matching genomes to spliced consensus sequence for 6 pangolin and 3 porcupine SRA datasets, plus Pangolin CoV GX-P4L. Reads were aligned to Pangolin CoV GX-P4L and bat-SL-CoVZC45 separately. A consensus from reads mapping to Pangolin CoV GX-P4L in the NSP4 C and N region was sliced with a consensus from reads mapping to bat-SL-CoVZC45 in the NSP10 and RdRp region. Empty gaps were removed (indicated by vertical lines, gap length as indicated). Note sequence coordinates are relative to the spliced consensus with non mapped nucleotide positions removed. Window size 5, step 5. Generated from Simplot generator hosted at http://babarlelephant.free-hoster.net/simplotSpikes.html

We ran a BLAST analysis against the nt database on a leading 25nt misaligned section of a 303nt contig covering the 3' end of the RdRp region. The 25nt sequence was found to have 100% identity (20/20nt, first 5nt not matched) to cloning vector pUC57, and no matches to adapter, coronavirus, or mammalian host mRNA sequences were found. The matching section of pUC57 is located in the multiple cloning site (MCS) just after an M13 forward sequencing primer (Supp. Fig. S77-78).

The design for the BioProject is specified as "TruSeq Stranded Total RNA Sample Preparation Kit". TruSeq stranded mRNA and stranded total RNA is strand specific and should only generate reads in a dominantly single direction (Illumina, 2017). We quantified the strandness for bat-SL-CoVZC45, and a strain of Human rubulavirus 2 (MH892406.1) in the 6 pangolin samples MJ-ZJ-MO-1/2/3/4/6 and MP-ZJ-MO-4 (Supp. Info. 6.4). First read forward (FR) to first read reverse (RF) ratios exhibited a strong bias in Human rubulavirus 2, moderate bias for Pneumonia virus of mice J3666 and only a slight bias in bat-SL-CoVZC45. The dual strandness of the bat-SL-CoVZC45 reads could indicate these reads may have been derived from dsDNA rather than cDNA, however a more rigorous analysis is warranted.

MEGAHIT was used for de novo assembly and contigs were aligned to bat-SL-CoVZC45 and pangolin CoV P4L using minimap2. A single 1,038nt covering a section of the ORF1a aligned to P4L was analyzed using blastn with a 94.6% identity (860/909nt) to 6 GX pangolin coronaviruses including PCoV_GX-P4L. A consensus sequence generated from MEGAHIT contigs was also analyzed using NCBI blastn with 1879/1981 (94.9%) and 792/801nt (98.9%) identity to Bat SARS-like coronavirus isolate bat-SL-CoVZC45 (Supp. Figs. S79-80).

Forty-one representative SARSr-CoV sequences from sequences producing significant alignments using blastn to the RdRp region of the novel CoV were downloaded multi-genome alignment. We used MUSCLE in MEGA11 (Tamura et al. 2021) to align the 42 betacoronavirus sequences, identified the RdRP coding region and extracted the nucleotide sequences for this region. Using the aligned RdRP coding sequences, we generated a maximum likelihood tree in MEGA11 using a general-time-reversible substitution model and 1000 bootstrap replicates (Fig. 13). The RdRp of the novel ZC45-related CoV is most closely related to bat-SL-CoVZC45 and SARS-related bat CoV Longquan-89 RdRps with ancestral lineage in common with SARS CoV strains, and markedly distant from the pangolin GX and GD CoVs.

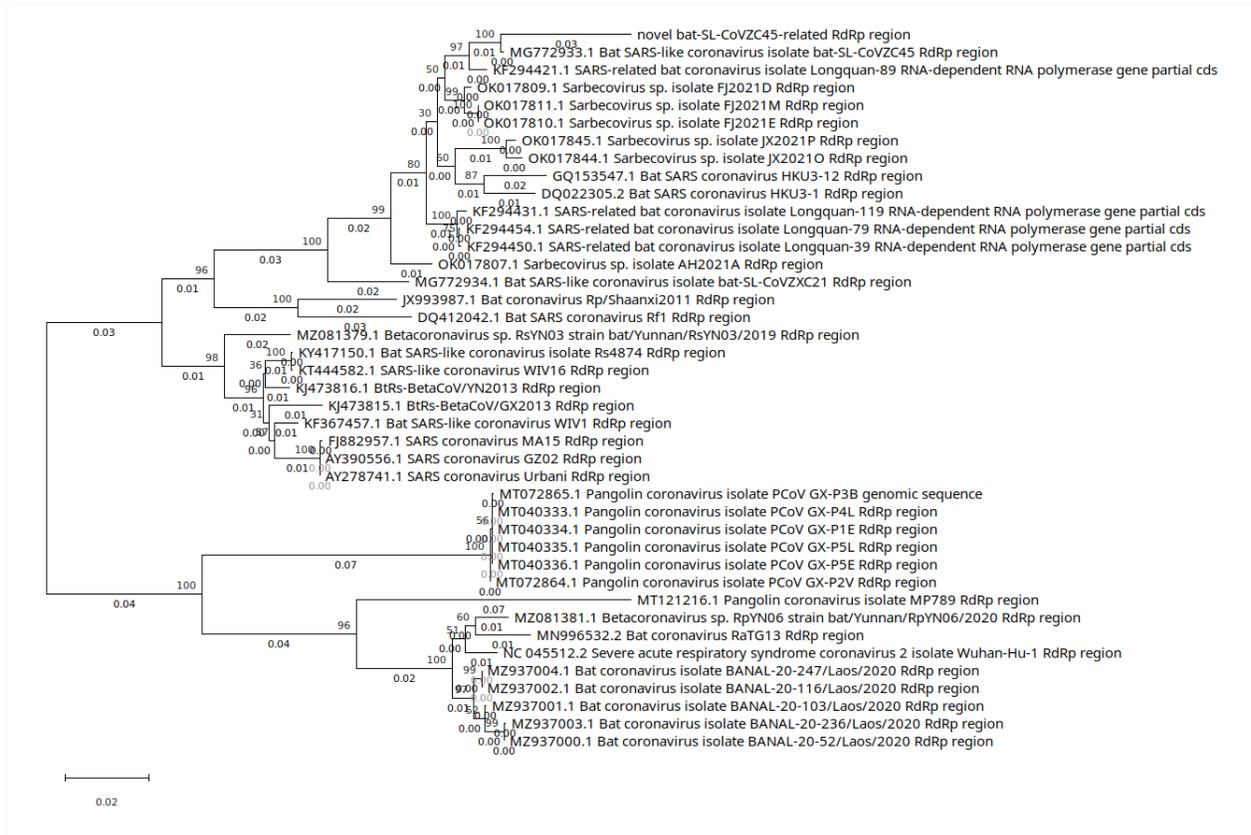

Fig. 13. Evolutionary history inferred using Maximum Likelihood and General Time Reversible model based on SARSr-CoV RdRp coding sequences. Tree drawn to scale, with branch lengths measured in the number of substitutions per site. Evolutionary analyses conducted and plotted using MEGA11 (Tamura et al. 2021). Novel bat-SL-CoV/ZC45-related RdRp sequence titled "novel bat-SL-CoV/ZC45-related RdRp region".

*Bacteria and primate sequences*

We tabulated a summary of NCBI STAT taxonomic analysis (Supp. Info. 6.0), and noted high bacterial levels for numerous non fecal samples. Datasets have an average of 42.5% bacteria for fecal samples, an average of 32.3% for nasal swab + anal swab +- throat swab +- fecal swab, and a bimodal distribution for mixed organ samples, with 7 samples with a <1% bacterial level (average 0.3%) as would be expected for organ samples, and 7 samples with a >5% bacterial percentage (average 26.3%). The 6 pangolin mixed organ samples with SARSr-CoV reads have bacteria percentages of 10.72% to 43.8% (average 29.73%), while the pangolin mixed organ samples without SARSr-CoV reads have bacteria percentages ranging from 0.02% to 5.46% (average 0.96%) with only a single sample (MJ-ZJ-MO-7) having bacteria levels >1%. High levels of primate sequence contamination was evident in multiple samples, with each of the 6 pangolin samples with SARSr-CoV reads containing between 3% and 30% primate sequences, which was further subclassified by STAT as dominantly Homininae subfamily. The 18 pangolin samples without SARSr-CoV reads had primate sequence contamination ranging from undetected to 2.09%. Notably the sample with the highest primate contamination was sample MJ-ZJ-MO-7, which of all the samples without SARSr-CoV reads, had the most anomalous bacteria content. Malayan porcupine nasal+anal swabs or nasal swab samples HB-FJ-NA-7, HB-HuB-N-3, HB-FJ-NA-3 were classified as having between 1.11%-7.52% Homininae origin content.

Significant human sequence contamination was also corroborated by allying a SerialAlign workflow to MJ-ZJ-MO-1/2/3 and MP-ZJ-MO-4 to sequentially filter out *Manis javanica*, then *Manis pentadactyla* matching reads with 100% identify, the remaining reads were mapped to *Chlorocebus sabaeus* (GCF_015252025.1) and separately to the *Homo sapiens* genome (GCF_000001405.39) both with 100% identity. While only trace Green monkey genome matches were found, 1-7% human genome read contamination was identified in each of the samples (Supp. Info. 6.5). Mitochondrial alignments also indicate significant human sequence content in the 6 pangolin samples and 3 *Hystrix brachyura* samples with SARSr-CoVs, ranging from 0.09% to 16.4% in pangolin samples to 4.8 to 76.7% in porcupine samples (Supp. Info. 6.7, 7.4). We note that *Rhinolophus ferrumequinum* mitochondrial reads were found in 5 of the 6 pangolin samples. We further note that no *Hystrix brachyura* mitochondrial sequences were found in any of MJ-ZJ-MO-1/2/3 and MP-ZJ-MO-4 pangolin samples and no *Manis javanica* or *Manis pentadactyla* mitochondrial genome sequences were found in SARSr-CoV containing *Hystrix brachyura* samples HB-HuB-N-3, HB-FJ-NA-3 and HB-FJ-NA-7 indicating that the SARSr-CoV reads were not associated with cross contamination by tissue samples from either of *Hystrix brachyura* or pangolin source.

Taxonomic classification of de novo assembled contigs from superorder Laurasiatheria classified at order level indicate MJ-ZJ-MO-1/2/4 to have high percentages of primate matching contigs (Supp. Fig. 81). Of the 104 de novo assembled contigs classified as rRNA using Metaxa2, multiple contigs of 100nt or greater in length which could be unambiguously identified and of animal origin, multiple were found to be from non pangolin origin (Supp. Info. 6.5, 6.6).

*Human orthorubulavirus 2*
All six pangolin samples containing bat-SL-CoVZC45-related CoV reads contained Human orthorubula virus 2, with MJ-ZJ-MO-4 and MJ-ZJ-MO-3 containing extremely high levels, at 5.91% and 27.66% of the metagenome respectively (Supp info. 6.1). To identify the closest strain, we aligned de novo assembled contigs in MJ-ZJ-MO-2 to the HPIV2 reference sequence and spliced two slightly overlapping contigs to generate a single sequence with near complete coverage of the genome and used blastn to identify Human orthorubulavirus 2 strain t146a293_HPIV2 (MH892406.1) as the closest match at 99.7% identity (15609/15649nt).

We found after de novo assembly of *Marmota himalayana* sample MH-HeB-NA-1 from PRJNA795267, 15 contigs which generated 62.31% coverage of the Human orthorubulavirus 2 t146a293_HPIV2 genome (MH892406.1). Using blastn to align the spliced contig sequence from MJ-ZJ-MO-2 to the consensus sequence from MH-HeB-NA-1 we found a 99.85% identity (12 single nt differences). Also using blastn, a 99.55% identity to Human orthorubulavirus 2 t146a293_HPIV2 (MH892406.1) was found for the MH-HeB-NA-1 consensus sequence. An NCBI STAT Krona taxonomic analysis of the nasal and anal swab sample MH-HeB-NA-1 showed the sample was comprised of 4% Hominoidea, while 0.06% of the metagenome was comprised of Rodentia, and 0.05% of *Marmota marmota marmota*. We further identified HPIV2 matching sequences in *Myocastor coypus* and *Paguma larvata* samples MC-FJ-NA-1, MC-HeB-T-2, PL-ZJ-NA-1 and PL-AH-MO-5, all with significant Catarrhini/Homininae/*Homo sapiens* content (Supp. Info. 7.0, 7.1). While He et al. (2022) interpret novel HIPV2 strains PPIV|Pangolin\China\ZJ-MO2\2019 and PPIV|Pangolin|China|ZJ-MO3|2019 to be pangolin hosted, the extremely high levels of HIPV2 seen in two pangolin samples and the presence of a highly similar strain in human sequence contaminated sample MH-HeB-NA-1 shows that the HPIV2 sequences are almost certainly of human origin and not a pangolin hosted virus. As a further check on STAT classification, we counted HPIV2 (MH892406.1) aligned reads in sample MJ-ZJ-MO-3 and found 30% of reads were HPIV2 sequences.

Other viruses with correlation to the presence of novel bat-SL-CoVZC45-related CoV reads in the six pangolin datasets are Pneumonia virus of mice J3666, several influenza strains and HIV-1 (Supp Figs. S7-8, S82-83, Supp. Info. 6.1).

**Avian feces or bat feces derived insect hosted densovirus**

BioProjects by Liu et al. (2019), Xiao et al. (2020), Li HM. et al. (2020) and He et al. (2022) all contained reads matching Parus major densovirus isolate PmDNV-JL with numerous SNV's. To further identify a potential closer virus sequence match, *de novo* assembled contigs from each SRA with significant PmDNV-JL reads matches were aligned to Parus major densovirus isolate PmDNV-JL (KU727766.1). Aligned contigs wered extracted and BLAST used against a local copy of nt database to identify closest matches (Supp. Info. 0.5). No systematic difference in contig matches to virus sequences was found between the BioProjects. 83% of matched contigs had highest homology to five avian hosted densovirinae all except Parus major densovirus isolate PmDNV-JL, submitted to NCBI by Xiao, Y. et al (2020) as part of an unpublished study. Yang et al. (2016) isolated Parus major Densovirus PmDNV-JL from Parus major lung tissue, but noted it "not possible to determine whether Parus major was infected by the densovirus or if the densovirus came from insects ingested by the bird without infection of avian cells". Prior to this study densovirinae had only been known to infect insects and Echinoderms, but not known to infect termites (Le Lay et al. 2020) or ants, the sole diet of *Manis Javanica* and dominant diet of *Manis pentadactyla*. Two of the matching genomes were isolated from bat feces (Ge et al. 2012) a third from bat organ tissues (Kohl et al. 2020, unpublished) and four from bird anal swabs. While desoviruses have been used as vector systems (Liu et al. 2016), we did not identify synthetic vectors attached the densivirus sequences in BioProjects reviewed here. We infer the most likely source of contamination is from bird feces and or bat feces material.

**Pangolin BioProject taxonomic and mitochondrion comparison**

A comparison of taxonomic classification, mitochondrial counts and bacteria percentages across datasets for 5 BioProjects (He et al. (2022) was studied separately) was made. Absolute mitochondrial counts and percentage coverage of 8 species found across the 5 BioProjects showed significant *Mus Musculus* and *Homo sapiens* mitochondrial contamination across all 5 BioProjects (Fig. 14; Supp. Fig. S85). Bacterial content by percentage calculated using NCBI STAT shows that of these 5 BioProjects, the Liu et al. (2019) SRA datasets have highest bacterial content (Fig. 15).

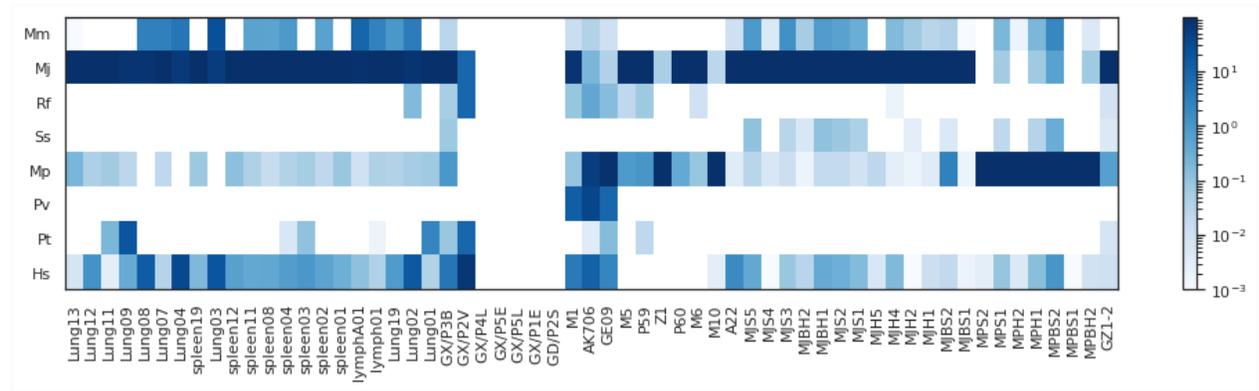

Fig. 14. Mitochondrial sequence percentages. Mm: *Mus musculus*, Mj: *Manis javanica*, Rf: *Rhinolophus ferrumequinum*, Ss: *Sus scrofa*, Mp: *Manis pentadactyla* , Pv: *Penaeus vannamei*, Pt: *Panthera tigris*, Hs: *Homo sapiens*. Log color scale.

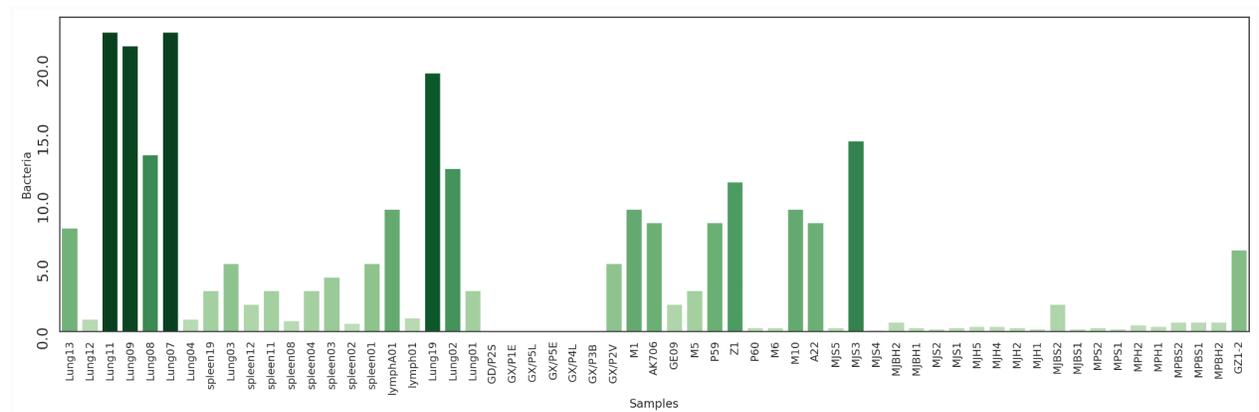

Fig. 15. Bacteria percentage in pangolin SRA datasets identified using NCBI STAT.

While taxonomic classification of contigs may not be truly representative of actual percentages of reads in datasets, the longer lengths of contigs allows higher confidence in classification. Using DIAMOND+MEGAN at order level for Chiroptera, Rodentia, Primates, Pholidota and Decapoda classification shows the highest levels of contigs classified as primates in Lung09, GX/P2V and Lung01, highest rodentia matches were in spleen11, spleen04 and MJS3 while highest chiroptera matches were classified in GX/P1E (Supp. Fig. S86-87). On a whole project level BioProject PRJNA573298 stands out as having an overall significant level of Primate and Rodentia matching contigs, while SRA's in PRJNA606875, PRJNA607174 and PRJNA610466 exhibit varying degrees of non-pangolin Animalia origin genomic contamination.

**Circular read pattern identified that is consistent with ligation-induced activities**

A circular read pattern of coverage of Pangolin CoV MP789 was identified in Xiao et al. (2020) PRJNA607174 sample M1. Reads upstream (towards 5' end) of 505nt and downstream of 1204nt consistently misaligned to Pangolin CoV MP789 (Supp. Fig. 88). We identified that when the reverse complement of the outlying sections of reads (423-504 nt and 1204-1284 nt) was taken, the resultant sequence then matched MP789 on the other side of this section of reads.

Similarly, in Liu et al. (2019) sample Lung11 had significant sections of misalignment that occured on 4 reads in the 16310-16694 nt region (Supp. Fig. S89). We found that for each of these reads (Supp. Info. 0.3 ids 1-4), when the reverse complement of the misaligned section was analyzed using BLASTN against the nt database, max score alignment was in each case Pangolin CoV MP789, with a maximum of one nt mismatch, and alignment location was in the same 16310-16694 nt region. Similar read end misalignment/alignment of reverse complement of misaligned part to the MP789 genome was seen for 2 reads from Lung09 (Supp. Info. 0.3 ids 28,29) and multiple reads in Lung08 (Supp. Fig. S90-91) and Lung07 (Supp. Fig. S92). Hassanin et al. (2020) interpreted that the circular read coverage was evidence for circular RNA molecules. However given the documented contamination and synthetic vectors in these datasets, and that circular RNAs are not generated by coronavirus RNA-dependent RNA polymerases, we interpret that the circular read patterns may be ligation-induced, but could potentially stem from artifacts from a sequencing platform different from the Illumina NovaSeq 6000 and Illumina Illumina HiSeq X Ten specified by the authors.

**Near-identical RBMs between SARS-CoV-2 and MP789**

It is still unclear how MP789 came to possess an RBM that shares 71 of 72 amino acids with SARS-CoV-2 and a perfect match with BANAL-20-103 and BANAL-20-52 (Supp. Fig. S93-94). Surprisingly, just 9 reads match the MP789 RBM in the entire PRJNA573298 BioProject (Liu et al. 2019): 2 reads from Lung07, 5 reads from Lung08, and 2 reads from Lung09 (Supp. Fig. S95).

Plots of a maximum likelihood tree of 18 SARSr-CoVs with closest RBD's to the MP789 RBD identified using blastn shows MP789 to cluster with SARS-CoV-2 and BANAL-20-52/103/236 (Supp. Fig. S96).

Although the RBMs of SARS-CoV-2 and MP789 share 71 of 72 amino acid residues, their underlying nucleotides differ more substantially: there are 24 differences in the 216 RBM nucleotides between them, all but one of which are silent mutations (SARS-CoV-2 RBM to MP789) (Supp. Fig. S97). While the MP789 RBM exhibits 23 nucleotide differences to the BANAL-20-103/BANAL-20-52 RBM regions (Supp. Fig. S98). Three of the 23 differences in the MP789 RBM to BANAL-20-103/BANAL-20-52 match the SARS-CoV-2 nucleotide sequence.

## Discussion

Two pangolin samples were found by Liu et al. (2019) to contain SARSr-CoV sequences, the first documented SARSr-CoV sequences in pangolin samples. Four other pangolin samples were later found to contain SARSr-CoV sequences (Liu et al. 2020; Lam et al. 2020, Xiao et al. 2020). In all 6 samples SARSr-CoV read counts are low, ranging from 984 to 4 reads (30 to 0.16 reads per million reads). In a study of game animals across China, He al. (2022) found no evidence for SARSr-CoV infection of any of the animals. However we identified six pangolin and three porcupine samples with SARSr-CoV reads, also with low read counts on the order of that seen for several samples sequenced by Liu et al. (2019). For pangolins with SARSr-CoV reads sequenced by Liu et al. (2019) and for datasets we found anomalously high bacteria levels for each of the samples containing SARSr-CoV sequences, much higher than would be expected from major organ samples. Additionally significant levels of human genome matching sequences and lesser levels of mouse genome origin reads were found in all samples.

BioProject PRJNA573298 (Liu et al. 2019) was a key dataset used for generating the pangolin CoV GD_1 (Xiao et al. 2020), MP789 (Liu et al. 2020) and Pangolin-CoV draft genome (Zhang, T. et al. 2020). Lung07, Lung08 were used by Liu et al. (2020) and Zhang, T. et al. (2020), while Xiao et al. (2020) additionally used Lung02, Lung11, Lung09 and Lung12 (Fig. 1). We make the following observations for SRA datasets in BioProject PRJNA573298:
1) The level of SARS-r CoV reads in PRJNA573298 are low, with the maximum MP789 read counts in Lung08. However the read count in this SRA dataset is 0.58X and 0.69X of the read counts for Parus major densovirus (or closely related avian feces or bat feces insect hosted densovirus) reads in Lung19 and lymph01. While pangolin respirovirus isolate M5 virus read counts in sample Lung19 were 146x of the MP789 read counts in sample Lung08. We note that SARS-r CoV and MERS-r CoV contamination of agricultural plant sequencing datasets in China in 2017 and 2020 was documented by Zhang, D. et al. (2020), with SARS-r CoV read levels significantly higher than for Pangolin CoV MP789 in BioProject PRJNA573298. Indeed, in addition to Parus major densovirus (or closely related avian feces or bat feces insect hosted densovirus), median read levels of Mus musculus mobilized endogenous polytropic provirus , Murine type C retrovirus and Caribou feces-associated gemycircularvirus are similar to median levels of Pangolin CoV MP789 across the BioProject. We interpret that the low level of reads in sample Lung08 (and even lower in the sample with next highest level, Lung07) are at a level that is consistent with sample contamination rather than bona fide animal infection.

2) We show using numerous genome filtering, rRNA and taxonomic workflows applied to both reads and contigs, that PRJNA573298 is significantly contaminated with human and mouse genomic sequences. We further found a significant number of complete reads exactly matching human and mouse mitochondria, in addition to tiger mitochondria;, 75% and 77% of the human and murine mitochondrial genomes were covered by contigs from the dataset, respectively.

3) A potential purpose for the vectorized human OCT4 and human POU5F1, SOX2, and MYC, found in PRJNA573298 is for induced pluripotent stem cell (iPSC) growth and preparation (Takahashi and Yamanaka, 2006). Given the *Homo sapiens* and *Mus musculus* genomic contamination of the dataset, we question whether iPSCs were being prepared for the establishment of human lung xenograft mice in a project being sequenced by the GIABR.

4) It is intriguing that the 5 datasets in PRJNA573298 containing Pangolin CoV MP789 are in a set of 6 datasets with >=12% bacteria content, compared with 7.5% average for PRJNA573298 and 4.2% across 5 BioProjects (Fig. 15). Lung19 being the only dataset with anomalously high bacterial content not containing MP789. We note that in a study of contaminated COVID-19 patient samples sequenced by the WIV (Zhou et al. 2020), Quay et al. (2021a) and Quay et al. (2021b) identified one BALF sample with an anomalously high level of bacteria was the most heavily contaminated by synthetic plasmids containing the Influenza HA gene sequence, as well vectorized Nipah virus sequences. While an association of SARSr-CoV reads and high bacteria content is also apparent for mixed organ pangolin datasets in He et al. (2022) (Supp. Info. 6.0).

The first peer-reviewed journal paper to identify the high identity between the SARSr-CoV sequences in Liu et al. 2019 datasets and SARS-CoV-2 was Zhang T. et al. (2020). Lung08 and Lung07 samples were used by Zhang T. et al. to identify a 91.02% sequence identity of 'Pangolin-CoV' to SARS-CoV-2. No NGS data however was generated. Lam et al. (2020) was published shortly after and sequencing data for 6 Guangxi pangolin samples and one Guangdong pangolin scale were deposited on NCBI. The batch of Guangxi pangolins are the only known pangolins other than the single batch Guangdong Wildlife Rescue Center sourced pangolin samples sequenced by Liu et al. (2019), Xiao et al. (2020), Liu et al. (2020) and likely Li HM. et al. (2020), and an sVNT assay in Wacharapluesadee et al. (2021) to have been reported to be carriers of SARSr-CoVs. Lam et al. (2020) merged Guangdong pangolin scale dataset GD/P2S from BioProject PRJNA606875 with Lung12, Lung09, Lung08, Lung07 and Lung02 from BioProject PRJNA573298 to generate GD pangolin-CoV. However, dataset GD/P2S contained 83 reads 75nt long exactly matching, and 132 reads of 75-nt length with a single nucleotide difference to SARS-CoV-2 indicating a significant contamination issue with this dataset. Furthermore the dataset had an estimated 99.995% of the original reads filtered out and as such is not a true metagenomic sample that can be relied upon to verify the GD pangolin-CoV genome. We also note further contamination issues in datasets in BioProject PRJNA606875. Vero-E6 based sample GX/P2V, from which Pangolin CoV GX/2V was isolated, contained 20 reads of 70-95 nucleotide length exactly matching the SARS-CoV-2 genome. While in the only non heavily filtered dataset, GX/P3B, 4.3% of mitochondrial aligned reads matched *Homo sapiens*. For *de novo* assembled contigs for the same dataset, 3.4% were found to match Rodentia, 7.1% matched *Homo sapiens* and 14.7% matched *Chiroptera* sequences after DIAMOND+MEGAN taxonomic classification. Finally, Primate rRNA sequences were found in GX/P5L, GX/P5E, GX/P4L, GX/P3B and Rodentia rRNA in GX/P5E, GX/P4L sample datasets.

Xiao et al. (2020) was next to deposit SRA data on NCBI with AK706 and GE09 submitted on 18/2/2020 but deleted on 10/3/2020 (Suryanarayana, 2020). Five phases of data uploading (including amplicon data replacement) ensued between the 22/4/2020 and 16/6/2021. Although Pangolin CoV GD_1 was isolated in Vero-E6 cells by Xiao et al. (2020), we have significant concerns over the claim by Xiao et al. (2020) that the virus was Pangolin hosted. The amplicon dataset GD1 published by Xiao et al. 2020 contains pangolin CoV GD_1 (GISAID: EPI_ISL_410721) viral sequences embedded in synthetic vectors. This is most evident for reads aligned to the 3' and 5' ends of the pangolin CoV GD_1 genome where vector sequences are longest. Cloning vector pEASY-tub/hptII was identified as the vector used, with synthetic plasmids presumably sequenced rather than pangolin organ samples. No mention of the use of synthetic cloning or expression vectors was found in the methods section of Xiao et al. (2020) or in an author correction (Xiao et al. 2021). Although in Xiao et al. (2021) two extra datasets (A22, P59) were stated to have been used in addition to those in Xiao et al. (2020) Extended Data Table 3 to derive the pangolin CoV GD_1 genome, we find that the GD_1 genome still cannot be generated without the use of the synthetically constructed GD1 amplicon dataset. While this dataset could have been part of an amplification step for resolving the 5' and 3' ends of the genome, it is not clear why the full genome would be covered, furthermore there no mention of this is referred to in their method description. As such we do not believe that pangolin CoV GD_1 can be relied upon to represent a true genome of a naturally occurring coronavirus infecting pangolins.

In addition to synthetic vectors in amplicon dataset GD1, we identified significant contamination by non SARS-r CoV related synthetic vectors in PRJNA607174 with large African Swine Fever Virus and Porcine circovirus 2 contigs assembled. Further indication of synthetic biology experimentation come from the identification of a circular read pattern of coverage of Pangolin CoV MP789/GD_1, with both positive and negative strands at misaligned ends of reads across a 5 read wide zone in sample M1. We interpret this pattern to be consistent with circularized cDNA generated during the ligation step of molecular cloning rather than naturally occurring circular RNA molecules (Hassanin et al. 2020). While the most complete example occurs in sample M1, several other occurrences of a circular read pattern were also identified in BioProject

PRJNA573298. This indicates that synthetic data may be contained in raw WGS datasets in addition to the GD1 amplicons as described above. Given that the GD_1 genome cannot be assembled without synthetic data, the Xiao et al. (2020) paper and GD_1 genome sequence (GISAID: EPI_ISL_410721) should be retracted.

Further evidence for contamination in PRJNA607174 is seen with Wenzhou shrimp virus 8 sequences were identified in samples M1, AK706 and GE09 (Fig. 1,6), the discovery of Wenzhou shrimp virus 8 from one of 5 shrimp species including *Penaeus vannamei* (Shi et al. 2016) and the identification of *Penaeus vannamei* sequences in the same samples using multiple independent methods, we interpret high confidence on shrimp sequence contamination of these three pangolin samples. We note that samples M1, AK706 and GE09 were the only samples from PRJNA607174 to be sequenced using Illumina NovoSeq 6000 machine id 'A00129' and all were sequenced on the same run with the same flowcell id (Supp. Info. 0.2). We note that *Penaeus vannamei* is the most dominant shrimp species in mariculture in China with dominant regions for production in Guangdong, Guangxi and Hainan provinces (Chen and Xiong, 2018).

Although pangolin sequencing BioProject PRJNA610466 by Li HM. et al. (2020) did not contain Pangolin CoV MP789/GD_1 sequences, SRA datasets were sequenced at the GIABR on the same illumina machine ('A00184') as 9 samples from PRJNA573298, and as such the dataset can be viewed as a 'negative control' for virus sequence comparison. We note a much lower level of bacteria for SRA's in PRJNA610466 compared with PRJNA573298, with MJS3 the only dataset to contain >5% bacteria (Fig. 14). However we do note a significant level of contamination by *Mus musculus* and *Homo sapiens* mitochondrial sequences, with 85% coverage of the *Mus musculus* mitochondrion sequence found in 10 SRA's, while >79% genome coverage of the *Homo sapiens* mitochondrion found in 9 SRA's. Comparing virus content (Fig. 1) and coverage (Supp Fig. S2), the only viruses only present in or more abundant in PRJNA573298 compared with PRJNA610466 are Pangolin CoV MP789, Pangolin respirovirus M5 and Parus major densovirus PmDNV-JL (or closely related avian feces or bat feces derived insect hosted densovirus). Numerous murine hosted viruses have similar read counts and coverage. Parainfluenza virus 5, RD114 retrovirus, SV40 and WHV sequences are only present in or more abundant in PRJNA610466. Given the presence of Parus major densovirus isolate PmDNV-JL (or closely related avian feces or bat feces derived insect hosted densovirus), and the fair similarity in multiple virus counts and coverage in both BioProject PRJNA610466 and BioProject PRJNA573298, the relationship of samples from these two projects should be clarified by the authors. While the identification of an integration junction between a lentiviral vector and *Manis Pentadactyla* genomic DNA indicates synthetic biology experiments on pangolin samples may have been undertaken, it is not clear however if the presence of a lentiviral transfer vector and SV40 Large T antigen is contamination-related.

Datasets for sample GZ1-2 and amplicon dataset 'journal_ppat_1009664_s001' were published by Liu et al. (2021) after Chan and Zhan (2020) noted that a the third pangolin discussed in Liu et al. (2020) supporting the generation of the MP789 had not been published and genome coverage was incomplete using available datasets. However, even with the inclusion of the amplicon dataset and sample GZ1-2, a 3nt gap in coverage of the MP789 genome and SNVs for the two reads covering position 5731nt relative to the reference genome. We thus question how Liu et al. 2020 could have generated the MP789 using the four datasets documented in Liu et al. 2021. We further note that sample GZ1-2 provided by Liu et al. 2021 in support of MP789 genome generation, had a very low Pangolin CoV MP789 read count, and consisted of 16 reads repeated 2-44 times, with 6 SNV's and a 6nt insert relative to MP789. Given that MP789 could not be assembled using data provided by Liu et al. (2020) and Liu et al. (2021) we call for the retraction of Liu et al. 2020 and clarification of the process used to generate the MP789 genome.

Two scenarios could explain the occurrence of bat-SL-CoVZC45-related sequences in the 6 Zhejiang pangolin samples sequenced by He et al. (2022): a natural zoonotic infection; or laboratory contamination. Bat-SL-CoVZC45 was isolated from a *Rhinolophus pusillus* bat sample in February 2017 in Zhoushan city, Zhejiang province, China (Hu et al. 2018). That all pangolin samples obtained and sequenced by He et al. were sourced from Zhejiang province, makes a natural infection seem plausible. However multiple factors indicate a natural infection is highly unlikely. He et al. note that "no viruses closely related to either SARS-CoV or SARS-CoV-2 (or other sarbecoviruses) were detected in any of the animals examined". As well as the 6 pangolins, 3 *Hystrix brachyura* from Fujian and Hubei provinces also had SARSr-CoV reads that we interpret likely to be sourced from the same single viral strain; *Hystrix brachyura* sample HB-FJ-NA-7 reads aligned to bat-SL-CoVZC45 had exactly the same SNV's as for the pangolin datasets in the NSP10 and mapped section of the RdRp coding region; Pangolin datasets MJ-ZJ-MO-3, MP-ZJ-MO-4 and porcupine datasets HB-FJ-NA-7 and HB-FJ-NA-3 each had best mapping of reads to bat-SL-CoVZC45 in the NSP10 and RdRp region and pangolin CoV GX/P4L in the NSP4 C and N termini regions; SARSr-CoV mapped reads exhibited a similar localized read distribution across all 6 pangolin and 3 porcupine datases; A match of a short unaligned section of a contig just upstream of the NSP10 region was found to match part of a cloning vector MCS just after the M13 FWD primer, and included the same *XbaI*, *EcoRV* and *BamHI* sites; A marked truncation at the 3' end with with reads in MJ-ZJ-MO-1, MJ-ZJ-MO-2 and MP-ZJ-MO-4 all exactly ending at 16239nt; and the dual strandedness of the SARSr-CoV reads obtained from a TruSeq Stranded Total RNA library. Taken together, we interpret the presence of a laboratory cultured and PCR amplified virus rather than natural infection of the animals. We caution against researchers placing a high confidence in the generated consensus sequences and contigs as being the exact viral strain(s) sequenced.

While He et al. (2022) claim the HPIV-2 sequences in the dataset are evidence for a novel pangolin Orthorubulavirus (PPIV|Pangolin|China|ZJ-MO2|2019 and

PPIV|Pangolin|China|ZJ-MO3-1|2019), that the metagenomic dataset for mixed organ sample MJ-ZJ-MO-2 comprises 46% bacteria, 26% HPIV-2, 13% Hominoidea and 2% Manis Javanica (using NCBI STAT Krona analysis) clearly shows an extensive contamination issue during preparation stages prior to sequencing, and given the higher human content than pangolin, we interpret the HPIV-2 virus reads to be human derived and not a novel pangolin hosted virus (He et al. 2022). The presence of matches to human influenza virus strains in 5 pangolin samples all of which contain SARSr-CoV reads, and >3.2% Primate read matching content could indicate contamination by virus infected human cell culture(s). Of note HPIV-2 was not detected in the 3 Malayan porcupine samples containing SARSr-CoV sequences, but was identified in 2 *Myocastor coypus*, 2 *Paguma larvata* and one *Marmota himalayana* sample, all with >4% Catarrhini/Homininae or *Homo sapiens* origin reads. In all, in addition to the points discussed above the anomalously high bacteria content for mixed organ samples (minimum of 10.7%) in the 6 pangolin samples, and the presences of reads of Homininiae origin (19.24% to 1.11%) in all nine pangolin and Malayan porcupine samples further support a laboratory contamination source for the SARSr-CoV sequences. On a further note, although all pangolin samples were collected from Zhejiang province, curiously sample MP-ZJ-MO-5 was sequenced on an Illumina machine with the same id ('A00184', run 827) as a machine at the GIABR used for sequencing pangolin samples from BioProjects PRJNA573298 and PRJNA610466.

Pangolin CoV (both MP789 and GD_1) binding to pangolin ACE2 is tenfold weaker than binding to human ACE2 (Wrobel et al. 2021). Furthermore, pangolin CoVs (both MP789 and GD_1) exhibit comparable binding to hACE2 as SARS-CoV-2, yet no outbreaks of human infection by pangolin CoVs was reported among pangolin handlers (Choo et al. 2020). Both of which are consistent with the interpretation here that the reads in BioProject PRJNA573298 (Liu et al. 2019) stem from contamination by hACE2-adapted SARS-r CoV experiments or culture of SARS-r CoV(s) with a SARS-CoV-2-like RBD using human cell lines, rather than from an evolutionary adaptation of a zoonotic virus to pangolins. We note sampling trips to Yunnan Province for bat sample collection by GIABR researchers and virus isolation in collaboration with WIV researchers (Tan et al. 2016, Tan et al. 2017), and further note bat sample collection approximately 4km from the location where bat CoV RmYN02 was sampled (Liang et al. 2020; Zhou et al. 2020), the Mojiang county (Tan et al. 2017) and near the border with Laos (Yang et al. 2019). GIABR researchers had also been involved in wide-scale sampling of bat coronaviruses across China (Latinne et al. 2020). All of which indicate the possibility for bat SARSr-CoV(s) to have been at the GIABR sequencing facility at the time of pangolin sequencing in 2019.

In a longitudinal survey of 334 pangolins in Malaysia (Lee et al. 2020) failed to find any evidence of coronaviruses or other potentially zoonotic viruses in pangolins in the wild. Similarly, in a metagenomic study, Hu et al. (2020) did not find any SARS-CoV-2 related reads in samples from one Malayan pangolin collected in 2017. Furthermore, Zhang D. (2020) did not find any reads matching pangolin CoV sequences in 93 pangolin samples from BioProject PRJNA529540 sequenced by Hu et al. (2020). Here we failed to find any trace of coronavirus sequences from 7 *Manis javanica* pangolins collected from the Guangdong Provincial Wildlife Rescue Center in BioProject PRJNA610466 (Li H. et al. 2020). We also did not find any SARSr-CoV matches using NCBI STAT and fastv microbial profile analysis of the 3 datasets from Guangxi sampled pangolins by Wenzhou University (PRJNA749865). Although we detected SARSr-CoV sequences in 6 of 24 pangolin samples sequenced by He et al. (2022), the authors did not detect any SARSr-CoVs in any of the animals studied and we interpret that the SARSr-CoV matching reads are due to laboratory contamination. Wacharapluesadee et al. (2021) found that a sample from one of ten pangolins confiscated from illegal traders in Thailand exhibited a strong SARS-CoV-2 sVNT inhibition result, and that they had access to an undocumeted study in China showing one of seven sampled pangolin sera at the Guangdong Wildlife Conservation and Protection Center in May 2003 tested positive for SARS-CoV using competition ELISA and VNT. The authors then mention that PCR testing using pan-CoV primers for the ten Thailand samples and the seven Guangdong samples from 2003 all tested negative. We note, in humans a recent history of infection by endemic human coronaviruses showed ELISA detection of SARS-CoV-2 S and N IgG antibodies and VNT of c. 10% (Ng et al. 2020) and an exposure of the pangolins to human HCoV-HKU1 or HCoV-OC43 is possible during smuggling. As the pangolin sera showed a negative PCR result a more detailed analysis is required to confirm a true bat derived zoonotic SARSr-CoV infection of the single Thailand 2020 captured and Guangdong 2003 captured pangolins. We further note Wacharapluesadee et al. do not clarify how the pangolin serum from the samples tested in 2003 were accessed for PCR testing with pan-CoV primers, nor why SARS-CoV-2 sVNT was not conducted on any preserved samples. Furthermore VNT titer test methods for Guangdong 2003 pangolins were not provided.

While Wenzel (2020) and Choo et al. (2020) propose humans may have been the source of the infection by pangolin CoV (GD_1/MP789), the 90.32% (Liu et al. 2020) of the MP789 genome/91.02% (Xiao et al. 2020) similarity of the GD_1 genome to SARS-CoV-2 is sufficiently different that pangolin CoV MP789 could not have evolved during the pangolins time in captivity. Given the contamination and synthetic vectors identified in the pangolin datasets discussed here, we believe the most parsimonious explanation for the occurrence of SARS-r CoV reads in pangolin datasets is that they were derived from contamination from hACE2 adapted SARS-r CoV experimentation, either from from a hACE2 adapted CoV, or a BANAL CoV relative sample. That pangolin CoV MP789/GD_1 has been widely used for SARS-CoV-2 origins studies is concerning and we urge all published studies relying on these pangolin CoVs to review their work in light of our findings.

## Conclusion

We have analyzed 6 pangolin metagenomic BioProjects, 5 containing sarbecoviruses, and discovered a novel SARSr-CoV in a dataset sequenced by He et al. (2022). We posit that the most parsimonious scenario for the occurrence of SARSr-CoVs in datasets by Liu et al. (2019) and He et al. (2022) is inadvertent contamination of the sequencing datasets. While we interpret Liu et al. to have mistakenly attributed the presence of SARSr-CoV reads to a genuine infection, He et al. failed to detect the presence of SARSr-CoV reads at all. For the BioProjects by Lam et al. (2020), Xiao et al. (2020) and Liu et al. (2020), while we cannot conclusively prove that SARSr-CoV reads did not stem from infected pangolins or accidental contamination, we make several observations that lead us to question the source of the SARS-CoV reads.

Lam et al. (2020), Xiao et al. (2020) and Liu et al. (2020) used pangolin samples from the same batch of smuggled pangolins as Liu et al. (2019) with all three papers having been submitted for publication within 9 days of each other. Curiously, using machines with similar sensitivity, Lam (2020) identified 2.6 times more SARSr-CoV reads from a keratinous pangolin scale as from the Liu et al. (2019) lung sample with highest SARSr-CoV read count, completely filtered out the non-SARSr-CoV reads, and used two separate machines for sequencing, combining the results. Furthermore the discovery of a novel SARSr-CoV in pangolin and porcupine datasets sequenced by He et al. (2020) raise questions as to the provenance of the 6 pangolin GX CoVs themselves. Parts of the novel genome are most closely related to pangolin GX CoVs, however, rather than a natural infection of the 6 pangolins and 3 porcupines, we hypothesize that the novel SARSr-CoV was sequenced from cell culture with inadvertent contamination of the animal datasets.

After the lack of publication of a gap filling dataset was noted by Chan and Zhan (2020), Liu et al. (2021) published a correction to Liu et al. (2020), but the published amplicon dataset, in combination with other datasets used by Liu et al. (2020), is still insufficient to allow full assembly of the pangolin CoV MP789 genome. Finally, the GD1 sample sequenced by Xiao et al. (2020) consisted of a synthetic dataset containing vectorised SARSr-CoV sequences, without which the pangolin CoV GD_1 cannot be assembled.

We call on Lam et al. (2020), Xiao et al. (2020) and Liu et al. (2020) to provide full unfiltered datasets and transparent documentation of the entire sampling and sequencing process so that cell culture contamination, even if inadvertent, can be ruled out. We further call on Liu et al. (2019) to address our contamination concerns and finally ask He et al. (2022) to explain the presence of what appears to be laboratory cultured virus sequences in 9 mammalian datasets, publish the full sequence of the novel virus strain and clarify similarities we note to both pangolin GX P4L and bat-SL-CoVZC45.

We caution researchers against the attribution of pangolin CoV GD_1, pangolin CoV MP789 and GX pangolin CoVs to a pangolin host until all these issues have been addressed.

# Methods

*Quality filtering and adapter trimming*

Reads in SRA datasets in BioProject PRJNA573298, PRJNA606875, PRJNA607174, PRJNA610466, PRJNA686836 and PRJNA793740 were trimmed and filtered using fastp version 0.21.0 (Chen et al. 2018) with default settings. All subsequent alignment and de novo assembly was done using the fastp cleaned reads. For the 6 pangolin datasets with SARSr-CoV reads in PRJNA793740 (MJ-ZJ-MO-1/2/3/4, MJ-ZJ-MO-6 and MP-ZJ-MO-4) adapter sequences were still evident, and the raw fastq datasets were trimmed using TrimGalore (Krueger, 2021) version 0.6.7 using default settings, which defaults to Illumina standard adapter detection and removal.

*Fastv microbial analysis*

All SRA's in BioProjects PRJNA573298, PRJNA606875, PRJNA607174, PRJNA610466, PRJNA686836 and PRJNA793740 were analysed against the Opengene viral genome kmer collection 'microbial.kc.fasta.gz' (https://github.com/OpenGene/UniqueKMER) downloaded on the 7/3/2021, using fastv (Chen et al. 2020).

*Alignment to specific viral sequences*

Several virus gene sequences were identified using fastv to have significant abundance (>~100 k-mer hits) across SRA's in BioProjects PRJNA573298, PRJNA606875, PRJNA607174, PRJNA610466, PRJNA686836 and PRJNA793740. Fastp cleaned reads were then aligned to selected virus genomes using either bowtie2 version 2.4.2 (Langmead and Salzberg, 2012) using the "--very-sensitve" alignment parameter or as otherwise specified, bwa mem version 0.7.17 (Li, 2013) using default parameters, or bwa mem2 version 2.0 (Vasimuddin et al. 2019) using default parameters. For BioProject PRJNA686836, datasets documented in Liu et al. (2021) were aligned to the MP789 genome using the "-Y" parameter, and with minimap2 using the "-x sr" parameter. GATK version 4.1.9.0 (Van der Auwera & O'Connor, 2020) was used to sort then mark duplicates. Samtools version 1.11 (Li et al. 2009) was used to index the marked .bam files for viewing in IGV version 2.8.13 (Thorvaldsdóttir et al. 2013). Samtools idxstats and Bamdst (Shiquan, 2018) version 1.0.9 were used for generating mapping statistics.

*Soft-clipped read end BLAST analysis.*

For amplicon dataset GD1, reads were aligned with bwa-mem2 to the Pangolin CoV GD_1 genome (GISAID: EPI_ISL_410721) using the "-Y" parameter. We extracted the soft-clipped sections greater than 14 bases long at read ends for both 5' and 3' ends. We then used local BLAST to query each soft-clipped section against the nt database using percentage identity 95% and expect value of 0.001.

*De novo assembly*

Fastp cleaned reads from each whole-genome sequencing SRA in PRJNA573298, PRJNA606875, PRJNA607174, PRJNA610466, PRJNA686836 and PRJNA793740 were *de novo* assembled using MEGAHIT v1.2.9 (Li et al., 2015) with default parameters. Amplicon dataset SRR13053879 was *de novo* assembled using MEGAHIT with the following parameters: "--k-max 79 --k-step 10".

*Primer and vector identification*

Each *de novo* assembled contig in PRJNA573298, PRJNA606875, PRJNA607174, PRJNA610466, PRJNA686836 and PRJNA793740 was queried against common primer and promoter sequences to identify potential synthetic constructs.

*Adapter identification*

For amplicon dataset GD1 (SRR13053879) we searched for the first 12 or 13nt of the following adapter sequences in the file: Illumina: AGATCGGAAGAGC; Small RNA: TGGAATTCTCGG; Nextera: CTGTCTCTTATA. None were detected. We analyzed the file using FastQC v0.11.9 (https://www.bioinformatics.babraham.ac.uk/projects/fastqc/).

*Taxonomic analysis and species identification*

NCBI SRA Taxonomy Analysis Tool (STAT) (Katz et al. 2021) was used for taxonomic classification. STAT is a MinHash-based k-mer tool using a lowest common taxonomic node classification with benchmarks showing a very similar F1-score on test datasets to Kraken2 (Wood et al. 2019). Each SRA in PRJNA573298, PRJNA606875, PRJNA607174, PRJNA610466, PRJNA686836 and PRJNA793740 was analysed using the online NCBI STAT tool to identify the lowest common taxonomical node statistics for each dataset.

*DIAMOND+MEGAN workflow*

Using a DIAMOND+MEGAN workflow (Buchfink et al. 2021; Huson et al. 2016), for each SRA in PRJNA573298 we ran DIAMOND version 2.0.13 with default sensitivity and the output format option "-f 100". We then ran meganizer version 6.21.17 and used MEGAN version 6.21.17 for taxonomic plots. For all BioProjects except PRJNA607174 we ran the same workflow on contigs generated from each SRA using MEGAHIT, and used the DIAMOND parameter "--sensitive". For PRJNA607174, due to large de novo assembled contig file sizes and time constraints we were not able to run DIAMOND analysis on contigs from samples P59, Z1, P60, M6, M10. At a genus, species and subspecies taxonomic classification level DIAMOND+MEGAN has been shown to have competitive F1 score, precision, recall, and AUPR results compared with 10 other metagenomic classifiers (McIntyre et al. 2017).

*Contamination filtering*

Several methods were used to differentiate *Homo Sapiens* and *Mus musculus* origin contaminating reads from reads of *Manis Javanica* or *Manis Pentadactyla* origin. In a strict classification workflow of read classification we termed 'SerialAlign', we align to genomes in serial, keeping only unaligned reads to align to each subsequent genome. For PRJNA573298 bowtie2 was used to align fastp processed reads from each SRA to the *Manis javanica* reference genome YNU_ManJav_2.0 (GCF_014570535.1), only unaligned reads were then aligned to the *Manis pentadactyla* reference genome YNU_ManPten_2.0 (GCF_014570555.1), then only unaligned reads were then to the *Homo sapiens* reference genome GRCh38.p13 (GCF_000001405.39). The final stage of alignment was repeated with unaligned reads from step 2 (*Manis Pentadactyla* alignment), aligned to the *Mus musculus* reference genome GRCm39 (GCF_000001635.27). We then used a python workflow to compare read id's for the final stage 3 alignments (mouse or human alignment) and classified reads which were found to align to both human and mouse genomes as "ambiguous". The entire process was repeated twice, once using the "--score-min L,0,0" parameter for each alignment stage, where only reads with 100% identity to the reference genome sequence were assigned, and a second run using the "--very-sensitive" parameter.

A further analysis to identify the quantity of human and mouse genome contamination of reads in PRJNA573298 using several methods outlined by Jo et al (2019) to identify the impact of mouse contamination in human sample genomic profiling as well as two more strict alignment methods. Jo et al. (2019) found that Disambiguate and ConcatRef exhibited robust sensitivity, specificity and F-scores when applied to read filtering mouse gene contamination of human sequencing data. Here we used a modified 'Disambiguate' workflow, with cvbio (Valentine, 2020), a modified version of Disambiguate (Ahdesmäki et al. 2017) used instead of Disambiguate to allow disambiguation of mappings to more than two genomes simultaneously. We applied the workflow to both raw reads (after being processed using fastp) and de novo assembled contigs. We used the following reference genomes: human GCF_000001405.39_GRCh38.p13_genomic, mouse GCF_000001635.27_GRCm39_genomic and pangolin GCF_014570535.1_YNU_ManJav_2.0_genomic reference. Disambiguate uses read alignment quality to classify reads that can be mapped to both species (Ahdesmäki et al. 2017).

The ConcatRef method as proposed by Jo et al (2019) was applied to reads and contigs in PRJNA573298. The following four genomes: *Homo sapiens* reference genome GRCh38.p13 (GCF_000001405.39), *Mus musculus* reference genome GRCm39 (GCF_000001635.27), *Manis javanica* reference genome YNU_ManJav_2.0 (GCF_014570535.1) and *Manis pentadactyla* reference genome YNU_ManPten_2.0 (GCF_014570555.1) were concatenated, then raw reads (after being processed using fastp) or contigs were aligned using bowtie2 v2.4.2 using default parameters and again using the parameter '--very-sensitive'. Alignment statistics for each SRA raw read and contig dataset was then calculated using Samtools v1.11 flagstat and stats (Danecek et al. 2021). Plots of statistics results were then generated using Matplotlib v3.3.4 (Hunter, 2007).

We used Seal (Bushnell, 2021) to align contigs in PRJNA573298 to multiple genomes simultaneously. Seal gave comparable performance to other methods, but was significantly faster and less memory intensive than any other genome filtering method applied. We used Seal for multi-genome alignment filtering for each *de novo* assembled SRA in BioProjects PRJNA607174 and PRJNA606875.

*Magic-BLAST alignment*

Each *de novo* assembled contig in PRJNA573298 was analyzed using MagicBLAST by querying each contg against the a local copy of the nt database using default parameters. MagicBLAST results were then filtered before analysis. A minimum total match length filter of 100nt was applied, and exactly matched sections were filtered to cover a minimum of 95% of the query sequence. Reads or contigs which did not meet the filter criterion were discarded.

*rRNA identification*

For each SRA in each BioProject we used Metaxa2 (Bengtsson-Palme et al. 2015) to identify rRNA sequences from fastp processed reads. Using a similar methodology to Massey (2021), for PRJNA573298, PRJNA607174 and PRJNA686836 we *de novo* assembled the metaxa2 results from each SRA using MEGAHIT with default parameters. For PRJNA606875, only SRR11093270 and SRR11093271 had sufficient reads for assembly. For PRJNA686836, only 5 contigs were generated and we used web NCBI BLAST to identify the most homologous rRNA sequences. For assembled rRNA contigs from PRJNA573298, PRJNA607174, PRJNA610466 and PRJNA793740 and rRNA reads and contigs from PRJNA686836, we used local NcbiCommandlineBlast with E=0.05 and 80% minimum identity and returned the top 100 results. Where more than one result had the same score and alignment length we searched for the strings 'manis' and 'pangolin' and if found, recorded the match as the best result. If no match was returned, we selected a result at random, and noted that multiple matches were found. For matches with five or less equivalent matches, we recorded the full title, and manually analyzed matches to determine if matches were unambiguously to the same species. As unambiguous matches are only recorded for specific cases, unambiguous match counts are not representative of percentage distribution.

*Mitochondrial alignment*

SRA's in PRJNA573298, PRJNA606875, PRJNA607174, PRJNA686836 and PRJNA610466 were each aligned to multiple mitochondria reference sequences. Bowtie2 v2.4.2 was used with the parameter "--local --score-min L0,0". For all but the three datasets with average post fastp read length of less than 100nt (Supp. Info 0.2) alignmed reads were then filteretered to include only matches of 100nt length or greater. Alignment statistics were calculated using bamdst (Shiquan, 2018) and samtools (Danecek et al. 2021).

*Yamanaka factor identification*

All SRA's in PRJNA573298 were pooled annd aligned to the following concatenated reference sequences: human SOX2 (NM_003106.4), MYC (NM_001354870.1), KLF4 (NM_004235.6) and five transcript variants for POU5F1 (NM_002701.6, NM_203289.6, NM_001173531.3, NM_001285986.2 and NM_001285987.1) using bowtie2 with the '--very-sensitive' parameter. The resultant fastq file was then de novo assembled using MEGAHIT with default parameters.

*Local BLAST*

Local BLAST (Camacho et al. 2009) analysis was conducted using NcbiblastnCommandline in BioPython using expect value=0.001 and percentage identity of 0.95 or as specified.

*Minimap alignment*

Contigs in PRJNA573298, PRJNA607174 and PRJNA610466 were aligned to a database of viruses downloaded from the Genome Sequence Archive on the 1/3/2021 using minimap2 version 2.20 (Li H. 2018).

*He et al. (2022)*

We used fastq-dump --split-3 on each SRA dataset from PRJNA793740 and PRJNA795267 used for analysis to extract paired and unpaired reads.

We found that adapters had not been trimmed from bat-SL-CoVZC45 using a standard fastp cleaning workflow. We use TrimGalore to trim all SRA datasets analyzed from BioProjects PRJNA793740 and PRJNA795267. Fastv analysis as specified in methods above was used to identify significant viruses with >c.5% coverage in the 24 pangolin datasets in PRJNA793740. Selected viruses were then combined with a suite of SARSr-CoV's (including Pangolin CoV MP789 and GX pangolin CoV's) to generate a reference of 65 viruses for alignment of each SRA using bowties using the parameter "--very-sensitive". The virus selection process was repeated for rodent datasets in PRJNA795267.

For the 6 pangolin datasets and 3 rodent datasets with SARSr-CoV reads, we pooled forward and reverse paired end read seats separately, and also pooled unpaired reads. We aligned pooled paired end and single end reads separately to both bat-SL-CoVZC45 and pangolin CoV GX P4L using bwa-mem using default parameters. After alignment we used samclip --max 30 to remove reads with large soft slipped ends. Samtools was used for conversion between sam and bam and statistics analysis.

We used MEGAHIT with default settings to de novo assemble each SRA dataset, and pooled paired end and single end datasets. Minimap2 with default settings was used to align contigs to both bat-SL-CoVZC45 and pangolin CoV GX P4L.

Kallisto version 0.46.2 was used to identify strandness of SRA datasets by using quant with "--fr-stranded" and "--rf-stranded" parameters.

## Acknowledgements

We thank Francisco de Asis for discussions on dataset metadata, and Steven Massey for feedback which helped improve the manuscript.

## References

Ahdesmäki MJ, Gray SR, Johnson JH, Lai Z. Disambiguate: An open-source application for disambiguating two species in next generation sequencing data from grafted samples. F1000Research. 2017;5(0):1-11. doi:10.12688/f1000research.10082.1

Andersen KG, Rambaut A, Lipkin WI, Holmes EC, Garry RF. The proximal origin of SARS-CoV-2. Nat Med. 2020;26(4):450-452. doi:10.1038/s41591-020-0820-9

Bağcı C, Patz S, Huson DH. DIAMOND+MEGAN: Fast and Easy Taxonomic and Functional Analysis of Short and Long Microbiome Sequences. Curr Protoc. 2021;1(3). doi:10.1002/cpz1.59

Bai J, Lin H, Li H, et al. Cas12a-Based On-Site and Rapid Nucleic Acid Detection of African Swine Fever. Front Microbiol. 2019;10(December):1-9. doi:10.3389/fmicb.2019.02830

Bengtsson-Palme J, Hartmann M, Eriksson KM, et al. metaxa2: Improved identification and taxonomic classification of small and large subunit rRNA in metagenomic data. Mol Ecol Resour. 2015;15(6):1403-1414. doi:10.1111/1755-0998.12399

Boratyn GM, Thierry-Mieg J, Thierry-Mieg D, Busby B, Madden TL. Magic-BLAST, an accurate DNA and RNA-seq aligner for long and short reads. bioRxiv. Published online 2018:1-19. doi:10.1101/390013

Buchfink B, Reuter K, Drost HG. Sensitive protein alignments at tree-of-life scale using DIAMOND. Nat Methods. 2021;18(4):366-368. doi:10.1038/s41592-021-01101-x

Bushnell B. BBTools 2021. http://sourceforge.net/projects/bbmap/

Camacho C, Coulouris G, Avagyan V, et al. BLAST+: architecture and applications. BMC Bioinformatics. 2009;10(1):421. doi:10.1186/1471-2105-10-421

Chan YA, Zhan SH. Single source of pangolin CoVs with a near identical Spike RBD to SARS-CoV-2. Published online 2020:1-15. doi:10.1101/2020.07.07.184374

Chen X, Xiong J. Development of the culture of the white-legged shrimp, Penaeus vannamei. Aquac China Success Stories Mod Trends. 2018;(Cui 2014):378-392. doi:10.1002/9781119120759.ch4_2

Chen S, Zhou Y, Chen Y, Gu J. Fastp: An ultra-fast all-in-one FASTQ preprocessor. Bioinformatics. 2018;34(17):i884-i890. doi:10.1093/bioinformatics/bty560

Chen S, He C, Li Y, Li Z, Melançon CE. A computational toolset for rapid identification of SARS-CoV-2, other viruses and microorganisms from sequencing data. Brief Bioinform. 2020;00(August):1-12. doi:10.1093/bib/bbaa231

Choo SW, Zhou J, Tian X, et al. Are pangolins scapegoats of the COVID‑19 outbreak‑CoV transmission and pathology evidence? Conserv Lett. 2020;13(6):1-12. doi:10.1111/conl.12754

Csabai I, Solymosi N, Istv´ I, Csabai I. Host genomes for the unique SARS-CoV-2 variant leaked into Antarctic soil metagenomic sequencing data. Published online 2022. doi:10.21203/rs.3.rs-1330800/v1

Danecek P, Bonfield JK, Liddle J, et al. Twelve years of SAMtools and BCFtools. Gigascience. 2021;10(2):1-4. doi:10.1093/gigascience/giab008

Deigin Y, Segreto R. SARS-CoV-2′s claimed natural origin is undermined by issues with genome sequences of its relative strains: Coronavirus sequences RaTG13, MP789 and RmYN02 raise multiple questions to be critically addressed by the scientific community. BioEssays. 2021;43(7). doi:10.1002/bies.202100015

Gao W-H, Lin X-D, Chen Y-M, et al. Newly identified viral genomes in pangolins with fatal disease. Virus Evol. 2020;6(1). doi:10.1093/ve/veaa020

Ge X, Li Y, Yang X, et al. Metagenomic Analysis of Viruses from Bat Fecal Samples Reveals Many Novel Viruses in Insectivorous Bats in China. J Virol. 2012;86(8):4620-4630. doi:10.1128/jvi.06671-11

Hassanin A. The SARS-CoV-2-like virus found in captive pangolins from Guangdong should be better sequenced. Published online 2020:3-8.

Hassanin A, Jones H, Ropiquet A. SARS-CoV-2-like viruses from captive Guangdong pangolins generate circular RNAs. Nature. Published online 2020. https://hal.archives-ouvertes.fr/hal-02616966/

He W-T, Hou X, Zhao J, et al. Virome characterization of game animals in China reveals a spectrum of emerging pathogens. Cell. Published online 2022. doi:10.1016/j.cell.2022.02.014

Hernández BD, García C, González M, et al. Monoclonal and Polyclonal Antibodies as Biological Reagents for SARS-CoV-2 Diagnosis Through Nucleocapsid Protein Detection. BioProcess J, 2021; 20. doi:10.12665/J20OA.Hernandez

Hu D, Zhu C, Ai L, et al. Genomic characterization and infectivity of a novel SARS-like coronavirus in Chinese bats. Emerg Microbes Infect. 2018;7(1). doi:10.1038/s41426-018-0155-5

Hu JY, Hao ZQ, Frantz L, et al. Genomic consequences of population decline in critically endangered pangolins and their demographic histories. Natl Sci Rev. 2020;7(4):798-814. doi:10.1093/NSR/NWAA031

Huang Z, Gong L, Zheng Z, et al. GS-441524 inhibits African swine fever virus infection in vitro.  Antiviral Research, 2021, 191, 105081. doi: 10.1016/j.antiviral.2021.105081

Hunter JD. Matplotlib: A 2D Graphics Environment. Computing in Science & Engineering, vol. 9, no. 3, pp. 90-95, 2007. doi: 10.1109/MCSE.2007.55

Huson DH, Beier S, Flade I, et al. MEGAN Community Edition - Interactive Exploration and Analysis of Large-Scale Microbiome Sequencing Data. PLoS Comput Biol. 2016;12(6):1-12. doi:10.1371/journal.pcbi.1004957

Illumina. TruSeq Stranded Total RNA Reference Guide, 2017. https://support.illumina.com/content/dam/illumina-support/documents/documentation/chemistry_documentation/samplepreps_truseq/truseq-stranded-total-rna-workflow/truseq-stranded-total-rna-workflow-reference-1000000040499-00.pdf

Jo SY, Kim E, Kim S. Impact of mouse contamination in genomic profiling of patient-derived models and best practice for robust analysis. Genome Biol. 2019;20(1). doi:10.1186/s13059-019-1849-2

Katz KS, Shutov O, Lapoint R, Kimelman M, Brister JR, O'Sullivan C. STAT: a fast, scalable, MinHash-based k-mer tool to assess Sequence Read Archive next-generation sequence submissions. Genome Biol. 2021;22(1):1-15. doi:10.1186/s13059-021-02490-0

Krueger F. Trim Galore, 2021. https://www.bioinformatics.babraham.ac.uk/projects/trim_galore/

Lam, T. T. Y., Shum, M. H. H., Zhu, H. C., Tong, Y. G., Ni, X. B., Liao, Y. S., Wei, W., Cheung, W. Y. M., Li, W. J., Li, L. F., Leung, G. M., Holmes, E. C., Hu, Y. L., Guan, Y. (2020). Identification of 2019-nCoV related coronaviruses in malayan pangolins in southern China. BioRxiv. https://doi.org/10.1101/2020.02.13.945485

Langmead B, Salzberg SL. Fast gapped-read alignment with Bowtie 2. Nat Methods. 2012;9(4):357-359. doi:10.1038/nmeth.1923

Latinne A, Hu B, Olival KJ, et al. Origin and cross-species transmission of bat coronaviruses in China. *Nat Commun*. 2020;11(1). doi:10.1038/s41467-020-17687-3

Lee J, Hughes T, Lee M-H, et al. No Evidence of Coronaviruses or Other Potentially Zoonotic Viruses in Sunda pangolins (Manis javanica) Entering the Wildlife Trade via Malaysia. Ecohealth. 2020;17(3):406-418. doi:10.1007/s10393-020-01503-x

Le Lay C, Shi M, Buček A, Bourguignon T, Lo N, Holmes EC. Unmapped RNA virus diversity in termites and their symbionts. Viruses. 2020;12(10):1-21. doi:10.3390/v12101145

Li D, Liu CM, Luo R, Sadakane K, Lam TW. MEGAHIT: An ultra-fast single-node solution for large and complex metagenomics assembly via succinct de Bruijn graph. Bioinformatics. 2015;31(10):1674-1676. doi:10.1093/bioinformatics/btv033. The

Li H. 2013. bwa-mem. http://bio-bwa.sourceforge.net/bwa.shtml#13

Li H, Handsaker B, Wysoker A, et al. The Sequence Alignment/Map format and SAMtools. Bioinformatics. 2009;25(16):2078-2079. doi:10.1093/bioinformatics/btp352

Li H. Minimap2: Pairwise alignment for nucleotide sequences. Bioinformatics. 2018;34(18):3094-3100. doi:10.1093/bioinformatics/bty191

Li HM, Liu P, Zhang XJ, et al. Combined proteomics and transcriptomics reveal the genetic basis underlying the differentiation of skin appendages and immunity in pangolin. Sci Rep. 2020;10(1):1-13. doi:10.1038/s41598-020-71513-w

Li X, Xiao K, Chen X, et al. Pathogenicity, tissue tropism and potential vertical transmission of SARSr-CoV-2 in Malayan pangolins. 2020a. BioXriv 2020. doi:10.1101/2020.06.22.164442

Li X, Giorgi EE, Marichannegowda MH, et al. Emergence of SARS-CoV-2 through recombination and strong purifying selection. 2020b. Sci Adv. 2020;6(27):eabb9153. doi:10.1126/sciadv.abb9153

Liang J, He X, Peng X, Xie H, Zhang L. First record of existence of Rhinolophus malayanus (Chiroptera, Rhinolophidae) in China. *Mammalia*. 2020;84(4):362-365. doi:10.1515/mammalia-2019-0062

Liu P, Li X, Gu J, et al. Development of non-defective recombinant densovirus vectors for microRNA delivery in the invasive vector mosquito, Aedes albopictus. Sci Rep. 2016;6(February):1-13. doi:10.1038/srep20979

Liu P, Chen W, Chen JP. Viral metagenomics revealed sendai virus and coronavirus infection of malayan pangolins (manis javanica). Viruses. 2019;11(11). doi:10.3390/v11110979

Liu P, Jiang J-Z, Wan X-F, Hua Y, Li L., Zhou J, Wang, X, Hou, F, Chen, J, Zou, J, Chen, J. (2020). Are pangolins the intermediate host of the 2019 novel coronavirus (SARS-CoV-2)? PLOS Pathogens, 16 (5), e1008421. https://doi.org/10.1371/journal.ppat.1008421

Liu P, Jiang J-Z, Wan X-F, et al. Correction: Are pangolins the intermediate host of the 2019 novel coronavirus (SARS-CoV-2)? PLOS Pathog. 2021;17(6):e1009664. doi:10.1371/journal.ppat.1009664

Mahmood A, Ali S. Microbial and Viral Contamination of Animal and Stem Cell Cultures: Common Contaminants, Detection and Elimination. J Stem Cell Res Ther. 2017;2(5):1-360. doi:10.15406/jsrt.2017.02.00078

Makarenkov V, Mazoure B, Rabusseau G, Legendre P. Horizontal gene transfer and recombination analysis of SARS-CoV-2 genes helps discover its close relatives and shed light on its origin. BMC Ecol Evol. 2021;21(1):1-18. doi:10.1186/s12862-020-01732-2

Massey S. SARS-CoV-2's closest relative, RaTG13, was generated from a bat transcriptome not a fecal swab: implications for the origin of COVID-19. arXiv. Published online November 17, 2021. http://arxiv.org/abs/2111.09469

McIntyre ABR, Ounit R, Afshinnekoo E, et al. Comprehensive benchmarking and ensemble approaches for metagenomic classifiers. Genome Biol. 2017;18(1):1-19. doi:10.1186/s13059-017-1299-7

Ng KW, Faulkner N, Cornish GH, et al. Preexisting and de novo humoral immunity to SARS-CoV-2 in humans. Science (80- ). 2020;370(6522):1339-1343. doi:10.1126/science.abe1107

Niu S, Wang J, Bai B, et al. Molecular basis of cross‑species ACE2 interactions with SARS‑CoV‑2‑like viruses of pangolin origin. EMBO J. Published online 2021:1-12. doi:10.15252/embj.2021107786

Quay SC, Rahalkar M, Jones A, Bahulikar R. Contamination or Vaccine Research? RNA Sequencing data of early COVID-19 patient samples show abnormal presence of vectorized H7N9 hemagglutinin segment. 2021a;(July). doi:10.5281/zenodo.5067706

Quay SC, Zhang D, Jones A, Deigin Y. Nipah Virus Vector Sequences in COVID-19 Patient Samples Sequenced by the Wuhan Institute of Virology. ArXiv: 2109.09112

Shi M, Lin XD, Tian JH, et al. Redefining the invertebrate RNA virosphere. Nature. 2016;540(7634):539-543. doi:10.1038/nature20167

Shiquan. Bamdst, 2018. https://github.com/shiquan/bamdst

Sims D, Sudbery I, Ilott NE, Heger A, Ponting CP. Sequencing depth and coverage: Key considerations in genomic analyses. Nat Rev Genet. 2014;15(2):121-132. doi:10.1038/nrg3642

Suryanarayana S. Altered datasets raise more questions about reliability of key studies on coronavirus origins. Dec. 29, 2020. https://usrtk.org/tag/pangolin-papers/

Takahashi K, Yamanaka S. Induction of Pluripotent Stem Cells from Mouse Embryonic and Adult Fibroblast Cultures by Defined Factors. Cell. 2006;126(4):663-676. doi:10.1016/j.cell.2006.07.024

Tamura K, Stecher G, Kumar S. MEGA11: Molecular Evolutionary Genetics Analysis Version 11. Mol Biol Evol. 2021;38(7):3022-3027. doi:10.1093/molbev/msab120

Tan B, Yang X Lou, Ge XY, et al. Novel bat adenoviruses with an extremely large E3 gene. *J Gen Virol*. 2016;97(7):1625-1635. doi:10.1099/jgv.0.000470

Tan B, Yang X Lou, Ge XY, et al. Novel bat adenoviruses with low G+C content shed new light on the evolution of adenoviruses. *J Gen Virol*. 2017;98(4):739-748. doi:10.1099/jgv.0.000739

Thorvaldsdóttir H, Robinson JT, Mesirov JP. Integrative Genomics Viewer (IGV): High-performance genomics data visualization and exploration. Brief Bioinform. 2013;14(2):178- 192. doi:10.1093/bib/bbs017

Valentine C. cvbio, 2020. https://github.com/clintval/cvbio#cvbio

Van der Auwera GA & O'Connor BD. (2020). Genomics in the Cloud: Using Docker, GATK, and WDL in Terra (1st Edition). O'Reilly Media.

Vasimuddin Md, Sanchit Misra, Heng Li, Srinivas Aluru. Efficient Architecture-Aware Acceleration of BWA-MEM for Multicore Systems. IEEE Parallel and Distributed Processing Symposium (IPDPS), 2019.

Wacharapluesadee S, Tan CW, Maneeorn P, et al. Evidence for SARS-CoV-2 related coronaviruses circulating in bats and pangolins in Southeast Asia. Nat Commun. 2021;12(1). doi:10.1038/s41467-021-21240-1

Wang X, Chen W, Xiang R, et al. Complete genome sequence of parainfluenza virus 5 (Piv5) from a sunda pangolin (Manis javanica) in China. J Wildl Dis. 2019;55(4):947-950. doi:10.7589/2018-09-211

Wei C, Lin Z, Dai A, et al. Emergence of a novel recombinant porcine circovirus type 2 in China: PCV2c and PCV2d recombinant. Transbound Emerg Dis. 2019;66(6):2496-2506. doi:10.1111/tbed.13307

Wenzel J. Origins of SARS‐CoV‐1 and SARS‐CoV‐2 are often poorly explored in leading publications. Cladistics. 2020;36(4):374-379. doi:10.1111/cla.12425

Wood DE, Lu J and Langmead B. Improved metagenomic analysis with Kraken 2. Genome Biol 20, 257 (2019). doi:10.1186/s13059-019-1891-0

Wrobel AG, Benton DJ, Xu P, et al. Structure and binding properties of Pangolin-CoV spike glycoprotein inform the evolution of SARS-CoV-2. Nat Commun. 2021;12(1):837. doi:10.1038/s41467-021-21006-9

Wu K, Liu J, Wang L, et al. Current state of global african swine fever vaccine development under the prevalence and transmission of ASF in China. Vaccines. 2020;8(3):1-26. doi:10.3390/vaccines8030531

Xiao K, Zhai J, Feng Y, et al. Isolation of SARS-CoV-2-related coronavirus from Malayan pangolins. Nature. 2020;583(7815):286-289. doi:10.1038/s41586-020-2313-x.

Xiao K, Zhai J, Feng Y, et al. Author Correction: Isolation of SARS-CoV-2-related coronavirus from Malayan pangolins (Nature, (2020), 583, 7815, (286-289), 10.1038/s41586-020-2313-x). Nature. 2021;600(7887):E8-E10. doi:10.1038/s41586-021-03838-z

Xiao Y, Shan T, Yang S and Zhang W. Viral genomes from wild and zoo birds in China. 2020. Unpublished. https://www.ncbi.nlm.nih.gov/nucleotide/MT138252.1?report=genbank&log$=nucltop&blast_rank=1&RID=0DT12VTJ01R

Yang et al. 2016. Genetic characterization of a densovirus isolated from great tit (Parus major) in China. doi:10.1016/j.meegid.2016.03.035

Yang R, Peng J, Zhai J, et al. Pathogenicity and transmissibility of a novel respirovirus isolated from a Malayan pangolin. J Gen Virol. 2021;102(4). doi:10.1099/jgv.0.001586

Yang WT, Shi SH, Jiang YL, et al. Genetic characterization of a densovirus isolated from great tit (Parus major) in China. Infect Genet Evol. 2016;41:107-112. doi:10.1016/j.meegid.2016.03.035

Yang X Lou, Tan CW, Anderson DE, et al. Characterization of a filovirus (Měnglà virus) from Rousettus bats in China. Nat Microbiol. 2019;4(3):390-395. doi:10.1038/s41564-018-0328-y

Young DF, Carlos TS, Hagmaier K, Fan L, Randall RE. AGS and other tissue culture cells can unknowingly be persistently infected with PIV5; A virus that blocks interferon signalling by degrading STAT1. Virology. 2007;365(1):238-240. doi:10.1016/j.virol.2007.03.061

Zhai SL, Chen SN, Xu ZH, et al. Porcine circovirus type 2 in China: An update on and insights to its prevalence and control. Virol J. 2014;11(1):1-13. doi:10.1186/1743-422X-11-88

Zhai SL, Zhou X, Lin T, et al. Reappearance of buffalo-origin-like porcine circovirus type 2 strains in swine herds in southern China. New Microbes New Infect. 2017;17:98-100. doi:10.1016/j.nmni.2017.02.009

Zhang T, Wu Q, Zhang Z. Probable Pangolin Origin of SARS-CoV-2 Associated with the COVID-19 Outbreak. Curr Biol. 2020;30(7):1346-1351.e2. doi:10.1016/j.cub.2020.03.022

Zhang, D. The Pan-SL-CoV/GD sequences may be from contamination. 2020. Zenodo doi: 10.5281/zenodo.5004213

Zhang D, Jones A, Deigin Y, Sirotkin K, Sousa A. Unexpected Novel Merbecovirus Discoveries in Agricultural Sequencing Datasets from Wuhan, China. 2021. arXiv:2104.01533 [q-bio.GN]

Zhou H, Chen X, Hu T, et al. A Novel Bat Coronavirus Closely Related to SARS-CoV-2 Contains Natural Insertions at the S1/S2 Cleavage Site of the Spike Protein. *Curr Biol*. 2020;30(11):2196-2203.e3. doi:10.1016/j.cub.2020.05.023

## Supplementary material

All supplementary material including the files listed below can be accessed at
doi: 10.5281/zenodo.6319735
Link: https://zenodo.org/record/6319735

**Supp. Figs.**

Supp_Figs.pdf

**Supp Info.**

0_Datasets.xlsx
1_PRJNA573298_analysis.xlsx
2_PRJNA606875_analysis.xlsx
3_PRJNA607174_analysis.xlsx
4_PRJNA610466_analysis.xlsx
5_PRJNA686836_analysis.xlsx
6_PRJNA793740_analysis.xlsx
7_PRJNA795267_analysis.xlsx

**Supp Data**

Porcine circovirus 2 contig: 3_2_SRR11119762_k141_294696.fa
ASFV contig: 3_3_SRR11119764_k141_1503693.fa
Coning vector: 3_1_SRR13053879_pEASY-T1_from_spliced_read_ids_21_25.fa
Genomic DNA junction: 1_2_SRR11306689_k141_9009

GD1 amplicon contigs K79_3, K79_5: 3_4_SRR13053879_k79_3_k79_5.fa
PRJNA573298 human iPSC aligned reads, de novo assembled: PRJNA573298_human_iPSC_aligned_MEGAHIT_final.contigs.fa
Novel SARSr-CoV from He et al. (2022) datasets reads aligned to bat-SL-CoVZC45 consensus: novel_ZC45-relatedCoV_ZC45_alignment_consensus.fa
Novel SARSr-CoV from He et al. (2022) datasets reads aligned to pangolin CoV GX_P4L consensus: novel_ZC45-relatedCoV_P4L_alignment_consensus.fa
Novel SARSr-CoV from He et al. (2022) datasets contigs aligned to pangolin CoV GX_P4L consensus: novel_ZC45-relatedCoV_P4L_MEGAHIT_pe_contigs_consensus.fa
Novel SARSr-CoV from He et al. (2022) datasets contigs aligned to bat-SL-CoVZC45 consensus: novel_ZC45-relatedCoV_ZC45_MEGAHIT_pe_contigs_consensus.fa

## Code

https://github.com/bioscienceresearch/Pangolin_CoV_SRA_analysis